\documentclass[10pt,journal,compsoc]{IEEEtran}
\usepackage{ifpdf}
\usepackage{multicol}
\usepackage{multirow}
\ifCLASSOPTIONcompsoc
  \usepackage[nocompress]{cite}
\else
  \usepackage{cite}
\fi
\ifCLASSINFOpdf
   \usepackage[pdftex]{graphicx}
\else
   \usepackage[dvips]{graphicx}
\fi
\usepackage{amsmath}
\usepackage{algorithmic}
\usepackage{array}
\ifCLASSOPTIONcompsoc
 \usepackage[caption=false,font=footnotesize,labelfont=sf,textfont=sf]{subfig}
\else
 \usepackage[caption=false,font=footnotesize]{subfig}
\fi

\usepackage{fixltx2e}
\usepackage{stfloats}
\usepackage{url}
\usepackage{color}
\usepackage{graphicx}
\usepackage{amsmath}
\usepackage{amssymb}
\usepackage{booktabs}
\usepackage{multicol}
\usepackage{multirow}
\usepackage{bbding}
\usepackage{url}
\usepackage{amssymb} 
\usepackage{mathrsfs}
\DeclareMathAlphabet{\mathpzc}{OT1}{pzc}{m}{it}
\usepackage{bbm}     
\usepackage{dsfont}  
\usepackage{yfonts}  
\begin{document}

\title{Learning to Immunize Images for Tamper Localization and Self-Recovery}

\author{Qichao Ying$^{\dagger}$,~\IEEEmembership{Student Member,~IEEE}, Hang Zhou$^{\dagger}$, Zhenxing Qian$^{\star}$,~\IEEEmembership{Member,~IEEE}, Sheng Li,~\IEEEmembership{Member,~IEEE} and Xinpeng Zhang,~\IEEEmembership{Member,~IEEE}

\IEEEcompsocitemizethanks{\IEEEcompsocthanksitem $^{\star}$ Corresponding author (E-mail: zxqian@fudan.edu.cn). $^{\dagger}$ The leading two authors contribute equally. Q.Ying, Z.Qian, S. Li and X. Zhang are with the School of Computer Science, Fudan University, Shanghai, China. H. Zhou is with the School of Computer Science, Simon Fraser University, British Columbia, Canada.
}
}

\markboth{Submitted to IEEE Transactions for review}
{Shell \MakeLowercase{\textit{et al.}}: Bare Demo of IEEEtran.cls for Computer Society Journals}

\definecolor{green}{rgb}{0, 0.5, 0}
\definecolor{orange}{rgb}{0.8, 0.6, 0.2}
\definecolor{orange2}{rgb}{1.0, 0.6, 0.2}
\definecolor{red}{rgb}{1.0, 0.0, 0.0}
\definecolor{teal}{rgb}{0.0, 0.4, 0.4}
\definecolor{purple}{rgb}{0.65,0,0.65}
\definecolor{saffron}{rgb}{0.95,0.75,0.2}
\definecolor{turquoise}{rgb}{0.0,0.5,0.5}
\definecolor{black}{rgb}{0.0, 0.0, 0.0}
\definecolor{gray}{rgb}{0.5, 0.5, 0.5}

\newcommand{\hang}[1]{{\color{black}[Hang: #1]}}
\newcommand{\hz}[1]{{\color{black}#1}}
\newcommand{\ying}[1]{{\color{black}#1}}
\newcommand{\yingmodification}[1]{{\color{black}#1}}
\newcommand{\redmarker}[1]{{\color{black}#1}}
\newcommand{\cg}[1]{{\color{black}#1}}

\newcommand{\minor}[1]{{\color{black}#1}}

\IEEEtitleabstractindextext{%
\begin{abstract}
Digital images are vulnerable to nefarious tampering attacks such as content addition or removal that severely alter the original meaning. It is somehow like a person without protection that is open to various kinds of viruses. Image immunization (Imuge) is a technology of protecting the images by introducing trivial perturbation, so that the protected images are immune to the viruses in that the tampered contents can be auto-recovered. This paper presents Imuge+, an enhanced scheme for image immunization. By observing the invertible relationship between image immunization and the corresponding self-recovery, we employ an invertible neural network to jointly learn image immunization and recovery respectively in the forward and backward pass. We also introduce an efficient attack layer that involves both malicious tamper and benign image post-processing, where a novel distillation-based JPEG simulator is proposed for improved JPEG robustness.  Our method achieves promising results in real-world tests where experiments show accurate tamper localization as well as high-fidelity content recovery. Additionally, we show superior performance on tamper localization compared to state-of-the-art schemes based on passive forensics.

\end{abstract}

}

\maketitle

\IEEEdisplaynontitleabstractindextext
\IEEEpeerreviewmaketitle

\IEEEraisesectionheading{\section{Introduction}\label{sec:introduction}}

\IEEEPARstart{D}{igital} images have largely replaced conventional photographs from all walks of life. \yingmodification{The Online Social Network (OSN) platforms like Instagram and Twitter are designed to amplify the power of image sharing, where users create, curate, and share unique images that spark conversation and speak for themselves.} However, digital images can hardly enjoy the credibility of their conventional counterparts.
The rapid advancements \redmarker{in} digital
image processing tools \redmarker{have} made it extremely easy to edit images for free, and the modified images can be shared in seconds with social networking services. 
Although most common image editing in life is benign and unprofitable, 
maliciously fabricated images can be utilized as a supplement to fake news or criminal investigation to potentially influence public opinion.
For example, a critical object can be replaced with an image patch from the same image, which is known as the \hz{\emph{copy-move attack}}.
Or a non-existing object can be added to the scene, which is known as the \hz{\emph{splicing attack}}
. \ying{Furthermore, image inpainting techniques~\cite{yu2019free,nazeri2019edgeconnect} have facilitated the removal of }
unwanted \hz{regions} without introducing noticeable artifacts.
What's worse, the readers are susceptible to well-crafted fake images and they might further circulate these fake images. 
In that sense, digital images are like \textit{people} without protection yet open to a variety of attacks in the wild, and the social networks will be paralyzed if crowded with \textit{sick people}.

\begin{figure*}[!t]
	\centering
	\includegraphics[width=1.0\textwidth]{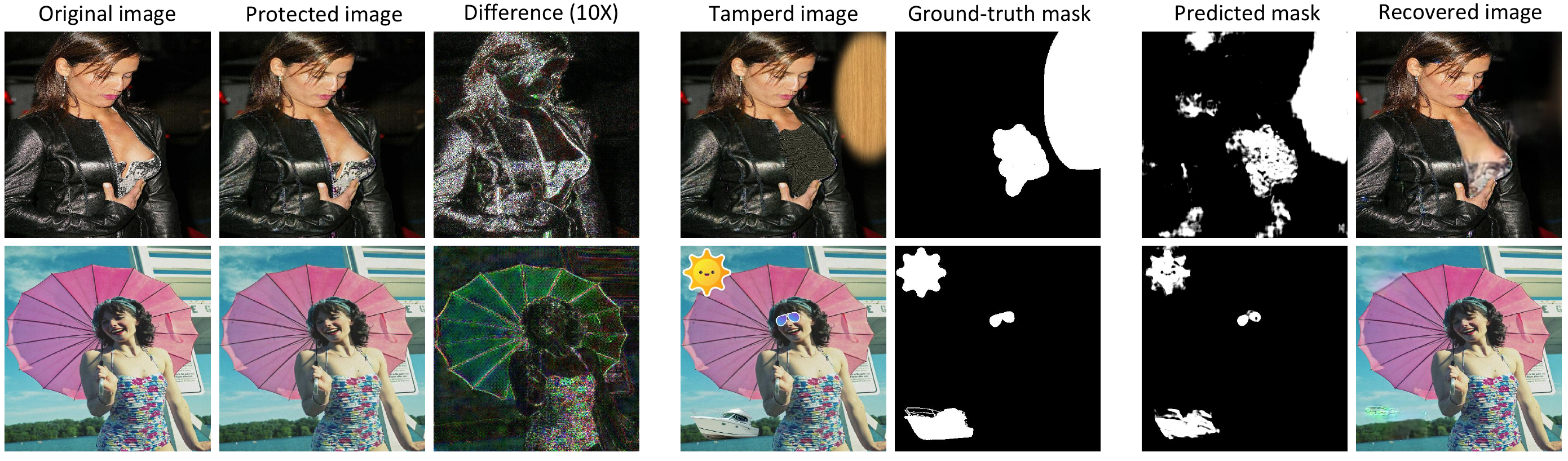}
	\caption{\textbf{Two examples of image tamper localization and content recovery by Imuge+.} Image immunization transforms the unprotected digital images into immunized versions where the residual is close to imperceptible. Malicious attacker deliberately tampers some of the contents and propagates the forged images in order to mislead the audience. Imuge+ localizes the tampered region and successfully recovers the images.}
	\label{img_teaser}
\end{figure*}

Image tamper localization has aroused extensive research interest to combat forged images and the security threats mentioned above.
The research of fake image forensics is to distinguish tampered images, or \textit{sick images}, from non-tampered images, or \textit{healthy images}. The technology is of considerable significance either for academia or industry.
With the emergence of \hz{deep learning}, the capability of image forgery detection is strengthened \hz{to a great extent}~\cite{wu2019mantra,islam2020doa,dong2021mvss}. 
\ying{These schemes are largely built upon detecting noise-level manipulation,} where the main component of the images is usually suppressed and the edge information as well as the noise distribution are studied for trace detection.
\hz{However,} the gains fail to meet the expectation in that it is hard to build a universal image tamper detection scheme, considering the enormous ways of image tampering \hz{in the} wild. In addition, these methods are often less effective on compressed or low-resolution images. 
Besides, none of the off-the-shelf image tamper detection methods \hz{are} equipped with self-recovery \hz{ability}. Once the images are \redmarker{manipulated}, it is hard for current techniques to reproduce the original contents. However, barely knowing the tampered regions \ying{cannot provide adequate information} for image forensics. It is usually hard to infer the intention of the attacks without the reference \redmarker{to} the ground-truth image, the intention of the editing, for example, how to distinguish benign image beautification from malicious attacks.

Image immunization is a novel technology \redmarker{for} protecting digital images by making them \textit{immune} to malicious attacks, or the \textit{viruses}. If the immunized images are \redmarker{manipulated} during image sharing on the OSNs, the tampered contents can be auto-recovered at the recipient's side without reference to any off-the-shelf image forgery detection or image reconstruction schemes. To enable immunization, the image owner is only required to embed some imperceptible perturbations into the image and uploads the immunized version instead of the original one. Since the embedding is trivial, the normal use of the targeted image is not affected, whereas the hidden information can resist common image processing attacks and help the recipient localize the tamper and recover the original image faithfully. Image immunization is originally proposed and implemented by a deep-network-based framework named Imuge in our conference paper~\cite{ying2021image} where we individually employ three networks for image self-embedding, tamper localization and image self-recovery. An attack layer is also proposed to simulate both tampering attacks and image benign processing attacks for robustness training.
Nevertheless, there are still many issues to be solved in Imuge. For example, when the critical components of an immunized image \redmarker{are} removed, only a blurry and approximate version of them can be reconstructed by Imuge. Besides, the localization accuracy is not satisfactory due to the poor generalization of robustness. Thirdly, Imuge cannot resist image inpainting attacks. It motivates us to develop an improved scheme for image immunization that can be applied on \redmarker{a} wilder range of images.


In this paper, we present an enhanced scheme denoted as Imuge+ for image immunization. We introduce trivial perturbation into the original image as immunization and blindly reconstruct the original contents in addition to tamper localization. The network is trained by multi-task learning and \hz{contains three modules}. The first \hz{one} is to keep the immunized image consistent with the original image, where the introduced perturbation is imperceptible. The second one is tamper localization \hz{by classifying} whether a pixel is altered or not. \hz{And} the third is \hz{image} self-recovery to encourage the recovered image \redmarker{to} resemble the original image.
Observing the invertible \ying{nature} between image immunization and self-recovery, we employ a normalizing-flow-based generator that jointly learns image immunization and self-recovery within a single network. The forward pass transforms an original image as well as its edge map into the corresponding immunized version. On receiving the attacked image, we use a localizer to determine the tampered areas by predicting the tamper mask, and in the backward pass of the generator, the hidden perturbation is transformed into information and we recover the original image as well as the edge map. Three most typical malicious attacks are simulated \hz{for effective network training}, e.g., copy-move and splicing, and benign attacks (rescaling, blurring). Furthermore, \ying{considering that most attackers deliberately hide their behavior by image filtering or compression, we also concatenate tamper simulation with image post-processing simulation,} where a novel knowledge-distillation-based JPEG generator (KD-JPEG) is developed for enhanced JPEG robustness. 



We showcase two examples of our scheme in Fig.~\redmarker{\ref{img_teaser}} where some crucial contents containing semantic information in the two protected images are removed by the malicious attacker. The recipient gets the JPEG-compressed tampered images and successfully retrieves the missing contents. Therefore, the recipient can identify the fake images and accordingly block their spreading.
We test our scheme by introducing \redmarker{human-participated} hybrid digital attacks, including copy-moving, splicing, inpainting and post-processing. The results demonstrate that our scheme can recover the tampered contents with high quality and fidelity. Compared with \ying{Imuge~\cite{ying2021image}, Imuge+ can combat more kinds of attacks such as copy-move and inpainting, where more details can be well preserved}. Additionally, we show the effectiveness of tamper detection of Imuge+ by comparison with some state-of-the-art \yingmodification{passive-based} image forensics schemes. 

This paper is an extended version of our conference paper~\cite{ying2021image}. We make the following new contributions:
\begin{enumerate}
\item We propose a \redmarker{tailored} network named Imuge+ for image immunization, \yingmodification{which reconstructs the original contents if the immunized images are manipulated.} We \yingmodification{develop the scheme} upon viewing the image immunization and recovery processes as a pair of inverse problem.
\item  We tempt to address the blurry result issue and the poor performance issue over JPEG compression by proposing a novel knowledge-distillation-based JPEG simulator as well as tamper\redmarker{ing}-based data augmentation. 
\item We have conducted comprehensive experiments to prove that \redmarker{our network design and training mechanisms} remarkably improves the overall performance of image immunization, both in the quality of the recovered image and the accuracy of \redmarker{image tampering} localization. 
\end{enumerate}


\section{Related Works}
\label{section_related_works}
In this section, we review the related works of Imuge+, namely, image protection using steganography, \redmarker{passive} image \redmarker{tampering} localization and image immunization.

\subsection{Image Protection using Steganography}
Image steganography aims at hiding secret information into the host images so that the recipient can extract the hidden message for covert communication.
In the past decades, many steganography-based schemes for image protection have been proposed, where the extracted secret information is utilized for image \redmarker{tampering} localization and fragile self-recovery.
For example, He et al.~\cite{he2006wavelet} and Zhang et al.~\cite{zhang2008fragile} respectively embed the Discrete Cosine Transform (DCT) coefficients and the Most Significant Bits (MSB), which is a compressed version of the image, into the Least Significant Bits (LSB).
Zhang et al.~\cite{zhang2009fragile} proposes to embed into the image blocks check-bits and reference-bits, where the former identif\redmarker{ies} the tampered blocks and the latter can exactly reconstruct the original image.
After that, Zhang et al.~\cite{zhang2010reference} proposes a reference sharing mechanism, in which the watermark to be embedded is a reference derived from the principal content in different regions and shared by these regions for content recovery. 
Later, Zhang et al.~\cite{zhang2011watermarking} proposes a watermarking scheme with flexible recovery quality. If the amount of extracted data in the areas without tampering is not enough, the method employs a compressive sensing technique to retrieve the coefficients by exploiting the sparseness in the DCT domain.
Besides, Korus et al.~\cite{korus2012efficient} theoretically analyzes the reconstruction performance with the use of communication theory.

Some works use steganography to prevent the images from being \redmarker{manipulated} by generative models.
Khachaturov et al.~\cite{khachaturov2021markpainting} proposes an adversarial method to attack inpainting systems by forcing them to work abnormally on the targeted images. 
Similarly, Yin et al~\cite{yin2018deep} proposes a defensive method based on data hiding to defeat Super-Resolution (SR) models.
The hidden information mainly resides in higher-band details of the targeted images, which are often analyzed or augmented by many schemes that employ deep networks for image restoration.

Though promising in the presented results, these methods are all fragile and typical image attacks \redmarker{on} the modified images can significantly weaken the performance. Besides, the steganography-based schemes against generative tampering are non-blind, where attackers can still \redmarker{modify} the protected images using traditional methods such as splicing and copy-move attack.

\subsection{\redmarker{Passive} Image Tamper\redmarker{ing} Localization}
\redmarker{Passive} image \redmarker{tampering} localization schemes aim at finding traces to unveil the behavior of image forgery.
Many existing image forensics schemes are specified on detecting typical kinds of attacks, which can be mainly categorized into three groups, i.e., splicing detection~\cite{kwon2021cat,salloum2018image}, copy-move detection~\cite{islam2020doa,li2018fast} and inpainting detection~\cite{zhu2018deep,li2019localization}.
\ying{As manipulating a specific region in a given image inevitably leaves traces between the tampered region and its surrounding, there are also many schemes for universal \redmarker{tampering} detection~\cite{dong2021mvss,wu2019mantra,hu2020span} that exploit such noise artifact.}
\ying{Li et al.~\cite{li2019localization} proposes to implement an FCN’s first convolutional layer with trainable high-pass filters and apply their HP-FCN for inpainting detection. 
Kown et al.~\cite{kwon2021cat} propose to model quantized DCT coefficient distribution to trace compression artifacts in splicing attacks.}
DOA-GAN~\cite{islam2020doa} proposes two attention modules for copy-move detection, where the first is from an affinity matrix based on the extracted feature vectors at every pixel, and the second is to further capture more precise patch inter-dependency. 

For universal \redmarker{tampering} detection, Mantra-Net~\cite{wu2019mantra} uses fully convolutional networks with BayarConv~\cite{bayar2018constrained} and SRMConv~\cite{srmconv} \ying{for feature extraction and further uses Z-Pooling layers as well as long short-term memory (LSTM) cells for pixel-wise anomaly detection}.
In MVSS-Net~\cite{dong2021mvss}, a system with multi-view feature learning and multi-scale supervision is developed to jointly exploit the noise view and the boundary artifact to learn manipulation detection features.
\ying{Hu et al.~\cite{hu2020span} proposes SPAN that models the relationship between image patches at multiple scales by constructing a pyramid of local self-attention blocks.}

Despite the existence of these well-designed works, real-world image \redmarker{tampering} localization is still an open issue. 
Besides, attackers can use a chain of image post-processing methods to hide their behaviors. Previous works are reported to have poor generalization against image post-processing or on JPEG-format images~\cite{wu2019mantra,islam2020doa}, \ying{where the learned clues can be easily erased.}
In this paper, we realize robust image \redmarker{tampering} localization using image self-embedding.

\subsection{\yingmodification{Image Restoration and} Immunization}
\yingmodification{Image restoration schemes reconstruct an image with higher visual quality.
For example, image inpainting schemes \cite{yu2018generative,nazeri2019edgeconnect} restore the contents within missing areas by referring to the ambient regions or the learned deep image prior~\cite{ulyanov2018deep}. Yu et al.~\cite{yu2018generative} proposes a CNN to synthesize novel image structures within the missing areas by explicitly utilizing surrounding image features as references.
EdgeConnect~\cite{nazeri2019edgeconnect} improves the visual quality of the generated images by reconstructing the edge information of the missing area ahead of image restoration.
LaMa~\cite{suvorov2022resolution} employs fast Fourier convolutions (FFC) to widen the receptive field to grasp more global statistical features for completing large missing areas.
However, the results of the inpainting schemes can be natural yet faulty compared to the ground truth in that providing visually pleasing results is the priority other than the reversibility. 
As a result, image inpainting is more often used to moderately manipulate an image for better layout rather than faithfully recover the image.

Compared to image restoration, image immunization~\cite{ying2021image} is a recently proposed novel technology that} protects images from being illegally tampered. The manipulated contents can be identified and auto-recovered at the recipient’s side without reference to sophisticated image forgery detection or image reconstruction schemes.
In~\cite{ying2021image}, a U-Net-based encoder~\cite{ronneberger2015u} is employed to conceal trivial information into the original image, where the hidden data serves as a \textit{vaccine} that helps conduct \redmarker{tampering} localization and image self-recovery. 
A differentiable attack layer is proposed to simulate both tampering attacks and image benign processing attacks for robustness training. We train a \redmarker{forgery detector} as well as an image decoder based on U-Net architecture. 
After the recipient gets the tampered protected image, he can get the original non-tampered version of the received attack image by \redmarker{tampering} localization and self-recovery.
Nevertheless, the scheme still \redmarker{has} several drawbacks. First, Imuge can only approximately recover the original contents within the tampered areas. There is still a big gap toward high-quality and accurate image recovery. Second, the accuracy of \redmarker{tampering} prediction is not satisfactory in many cases due to poor generalization. It motivates us to present an enhanced scheme for image immunization to comprehensively address the above issues.

\section{Method}
\label{section_methods}
\begin{figure*}[!t]
	\centering
	\includegraphics[width=1.0\textwidth]{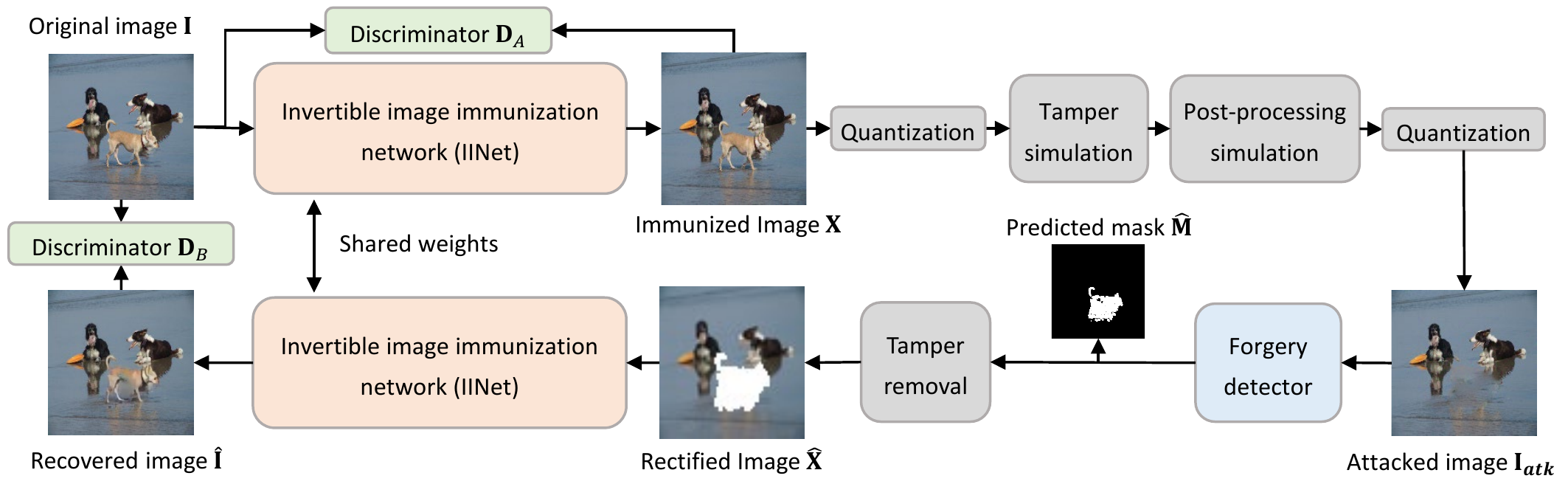}
	\caption{\textbf{Sketch of the pipeline of Imuge+.} IINet is employed to embed slight perturbations into the original images for immunization. Hand-crafted digital attacks in the social medias are decomposed into tampering and post-processing and correspondingly simulated by an attacking layer. Afterwards, a forgery detector helps predicting the tampering mask and removing the tamper from the received image. Finally, the inversed procedures of IINet recovers the original image. We additionally introduce two discriminators to improve the quality of the generated images.}
	\label{fig_framework}
\end{figure*}

In this section, we elaborate on our method for image immunization and self-recovery. We first present the problem statement of image immunization and self-recovery. Then, we show the details of the network architecture and the learning objectives. Finally, we introduce the training mechanisms.

\subsection{Approach Overview}
\noindent\textbf{Problem Statement. }
Typical malicious image manipulation attack can be generalized by a function $\emph{Mani}(\cdot)$ in Eq.~(\ref{eqn_tamper}). 
\begin{equation}
\label{eqn_tamper}
\mathbf{X}_{\emph{atk}}=\emph{Mani}(\mathbf{X},\mathbf{M})=\emph{IP}\left(\mathbf{X}\cdot(1-\mathbf{M})+\mathbf{R}\cdot\mathbf{M}\right),
\end{equation}
where $\cdot$ denotes pixel-wise multiplication. $\mathbf{X}$ and $\mathbf{X}_{\emph{atk}}$ are respectively the targeted image and its attacked version. $\mathbf{M}$ is the tampering mask, which is generally the region of interest of $\mathbf{X}$. The contents within the mask are replaced with irrelevant contents denoted by $\mathbf{R}$. $\emph{IP}(\cdot)$ summarizes the image post-processing behaviors such as JPEG compression, rescaling, etc. 
Image \redmarker{tampering} localization is to determine the \redmarker{tampering} mask $\mathbf{M}$ from the received attacked image $\mathbf{X}_{\emph{atk}}$, and image self-recovery is to take an advanced step to reconstruct the destroyed contents, i.e., $\mathbf{X}\cdot\mathbf{M}$. 
We employ two function\redmarker{s} $f_{M}(\cdot)$, $f_{\emph{rec}}(\cdot)$ respectively for these two tasks.
\begin{equation}
\label{equation_recovery}
\hat{\mathbf{X}}=f_{\emph{rec}}\left(\mathbf{X}_{\emph{atk}}\cdot\left(1-\hat{M}\right)\right),
\end{equation}
\begin{equation}
\label{equation_tamper_mask}
\hat{M}=f_{\emph{M}}(\mathbf{X}_{\emph{atk}}),
\end{equation}
where $\hat{\mathbf{X}}$ and $\hat{\mathbf{M}}$ are respectively the \redmarker{reconstructed image and the predicted tampering mask}.
However, since the non-tampered areas of $\mathbf{X}_{\emph{atk}}$ do not contain any information of $\mathbf{X}\cdot\mathbf{M}$. As a result, if $\mathbf{X}$ in Eq.~(\redmarker{\ref{eqn_tamper}}) is directly the original image $\mathbf{I}$, it leads to a sub-optimal solution that the destroyed contents will be hallucinated using image prior~\cite{ulyanov2018deep} rather than faithfully recovered.

To address the issue, we employ a third function $f_{prt}(\cdot)$ to embed deep representations of an original image into itself, i.e.,
$\mathbf{X}=f_{\emph{prt}}(\mathbf{I})$ in Eq.~(\ref{eqn_tamper}),
and formulate $\mathbf{X}=f_{\emph{prt}}\left(\mathbf{I}\right)$ and $\hat{\mathbf{I}}=f_{\emph{rec}}(\mathbf{X}_{\emph{atk}}\cdot\left(1-\hat{\mathbf{M}}\right))$ as a pair of invertible functions, where we expect that $\hat{\mathbf{I}}=\mathbf{I}$ and $\hat{\mathbf{M}}=\mathbf{M}$. In other words, by introducing image self-embedding, we wish to recover $\mathbf{I}$ when the tampered area is detected and removed, even if the self-embedded version $\mathbf{X}$ is attacked with randomized attacking method $\emph{IP}(\cdot)$ and \redmarker{tampering} mask $\mathbf{M}$.

\noindent\textbf{Network Modeling. }
Accordingly, we design four phases for Imuge+, namely, image immunization, image redistribution, forgery detection and image recovery. 
We use a single INN network called Invertible Immunization Network (IINet) to model $f_{\emph{prt}}(\cdot)$ and $f_{\emph{rec}}(\cdot)$ simultaneously. Image tampering and post-processing are implemented by an attacking simulation layer which consists of several differentiable methods. A forgery detector is trained for $f_{\emph{M}}(\cdot)$.
We also introduce two discriminators denoted as $\mathcal{D}_A$ and $\mathcal{D}_B$ to respectively distinguish $\mathbf{X}$ and $\hat{\mathbf{I}}$ from $\mathbf{I}$. 
Fig.~\ref{fig_framework} shows a sketch of the pipeline of our scheme. 

Specifically, we transform the original image into the immunized image using the forward pass of IINet. The protected version is uploaded onto the social cloud for daily applications instead of the unprotected original image. To simulate the image redistribution stage, we first perform tampering that evenly varies from splicing, copy-move or inpainting attack. Then, common image post-processing attacks are performed before data storage. On the recipient's side, the forgery detector produces the predicted \redmarker{tampering} mask to see which parts of the image are \redmarker{manipulated}, and we correspondingly remove the tampered contents. Finally, by inversely running the IINet, we reconstruct the recovered images. 
If the generated images are of high quality, they can evade the classification of the two discriminators.

\subsection{Framework Specification}

\noindent\textbf{Invertible Immunization Network (IINet)}
IINet embeds deep representation of the original image into itself, and conducts self-recovery provided with the rectified image. 
It is empirical that image immunization should not affect normal use of the image.
We accepts the invertible U-shaped network proposed by Zhao et al. \cite{zhao2021invertible}, where the involved wavelet-based blocks can provide empirical biases to decompose the generated features into lower and higher sub-bands. The design helps IINet restrict modifications towards lower sub-bands and conceal the information required by immunization in the higher sub-bands. 
We \yingmodification{additionally} generate the edge map $\mathbf{E}$ of the original image using the canny edge detection algorithm as the additional input of IINet. 
The introduction of the edge map is to aid high-fidelity image recovery, which is inspired by EdgeConnect~\cite{nazeri2019edgeconnect}.
The edge map is a one-channel matrix, and therefore the channel number of the input is four. Accordingly, there will be an additional output channel, denoted as $\mathbf{Y}$. We nullify $\mathbf{Y}$ by forcing it to be close to a zero matrix $\mathbf{O}$. In the inversed process, we also feed an extra $\mathbf{O}$ and let IINet output the predicted edge map $\hat{\mathbf{E}}$ along with $\hat{\mathbf{I}}$. 

\begin{figure}[!t]
	\centering
	\includegraphics[width=0.5\textwidth]{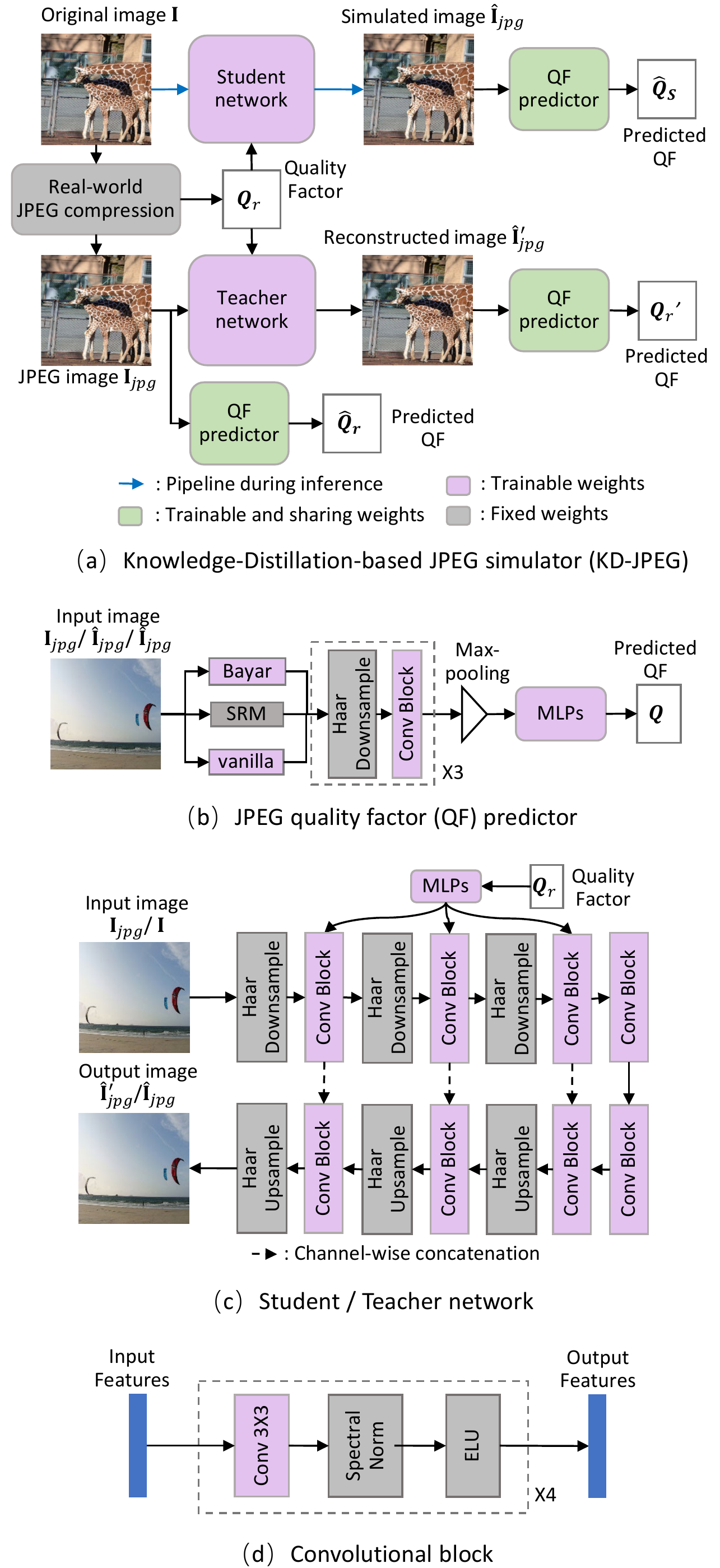}
	\caption{\textbf{Illustration of KD-JPEG.} The student/teacher network are respectively provided with the plain-text image and its JPEG version using real-world compressor. The two networks output the generated JPEG images. The predictor is pre-trained by classifying the real-world JPEG images and guides the generation of the simulated JPEG images. \yingmodification{To save computational complexity, these networks are based on basic architectures}. During inference, we only activate the student network.}
	\label{fig_jpeg}
\end{figure}
\noindent\textbf{Attack simulation.} 
We introduce the malicious attacks before introducing benign post-processing attacks.
To begin with, \redmarker{several} random areas within $\mathbf{X}$ are selected and \redmarker{manipulated} according to the \redmarker{tampering} mask $\mathbf{M}$. We denote the tampered image before image post-processing attack as $\mathbf{X}_{\emph{tmp}}$, where $\mathbf{X}_{\emph{tmp}}=\mathbf{X}\cdot(1-\mathbf{M})+\mathbf{R}\cdot\mathbf{M}$. The generation of $\mathbf{M}$ is borrowed from DeepFillV2~\cite{yu2018generative}. We restrict the rate of the tampered region as $r_{t}\in[0.0,0.5]$. To simulate \yingmodification{the} copy-move attack, we let $\mathbf{R}$ in Eq.~(\ref{eqn_tamper}) as a spatially-shifted version of $\mathbf{I}_{M}$, which also contains the embedded information. To simulate \yingmodification{the} splicing attack, we simply use another clean image for replacement. To simulate \yingmodification{the} inpainting attacking, we use the open-source model from DeepFillv2~\cite{yu2018generative} to generate the inpainted result as $\mathbf{R}$. The three kinds of tamper are iteratively and evenly performed for balanced training. 
Note that minimizing the false alarm rate is also important for computer vision tasks. Accordingly, we force Imuge+ to predict a zero matrix for \redmarker{tampering} localization when there is no tampering attack, i.e., $r_{t}=0$ and $\mathbf{X}_{\emph{tmp}}=\mathbf{X}$.

After image tampering, we implement the most-commonly-seen image post-processing attacks to simulate that the attacker wants to conceal the tampering behavior. $\mathbf{X}_{\emph{atk}}=\emph{IP}(\mathbf{X}_{\emph{tmp}})$. During training, we simulate the most common types of image post-processings, including: Addition of White Gaussian noise (AWGN), Gaussian blurring, image rescaling, lossy image compression (PNG and JPEG), median filtering and cropping. Note that if $\mathbf{X}$ is cropped, the recipient only conducts forgery detection and image recovery within the cropped area.

Finally, both $\mathbf{X}$ and $\mathbf{X_{\emph{atk}}}$ are required to be converted into 8-bit RGB format. We perform differentiable image quantization using Straight-Through Estimator~\cite{bengio2013estimating} for gradient approximation.

\noindent\textbf{Knowledge-distillation-based JPEG simulator.}
Fig.~\ref{fig_jpeg} illustrates the network design of \redmarker{KD-JPEG}, which consists of three parts, namely, a real-world JPEG compressor, a QF predictor, and a pair-wise student and teacher network.
\yingmodification{Though} there are already many schemes \redmarker{that} include a carefully-designed JPEG simulator, e.g., Diff-JPEG~\cite{shin2017jpeg}, MBRS~\cite{jia2021mbrs} and HiDDeN~\cite{zhu2018hidden}, the real-world JPEG robustness of these schemes is still reported to be limited. 
It mainly attributes to that neural networks can over-fit fixed compression mode where these works use \yingmodification{fixed codes and} limited number of quantization tables. In contrast, the quantization table in real-world JPEG images can be customized and \yingmodification{more} flexibly controlled by the Quality Factor (QF) as well as the image content. 
We present a novel knowledge-distillation-based JPEG simulator using a novel JPEG generative \redmarker{network} to better \yingmodification{approximate real-world JPEG compression}.

First, we use the FFJPEG library to store the plain-text images $\mathbf{I}$ in JPEG format as the ground-truth JPEG-compressed images $\mathbf{I}_{\emph{jpg}}$, with ground-truth QF $Q_r$ arbitrarily assigned. 
We then train the QF predictor to classify $\mathbf{I}_{\emph{jpg}}$ by their QF. 
\yingmodification{Here we empirically set the labels of the QF classification task as ${C}_{\emph{QF}}=\{10,30,50,70,90,100\}$, where ``100" represents ``not compressed". We do not introduce more labels in between such as $\{20,40,...,80,...\}$ in that when an image is compressed using two close QF ($\delta_{\emph{QF}}<10$), the discrepancy is not significant enough for classification. We allow the network to give an imprecise prediction when $Q_r\notin{C}_{\emph{QF}}$.}
Afterwards, we train the pair-wise teacher and student network to produce pseudo-JPEG images.
The student network is provided with the plain-text image $\mathbf{I}$ while the teacher network is given with the corresponding real-world compressed image $\mathbf{I}_{\emph{jpg}}$. 
They are to produce $\hat{\mathbf{I}}_{\emph{jpg}}$ and $\hat{\mathbf{I}}_{\emph{jpg}}^{'}$, which should resemble $\mathbf{I}_{\emph{jpg}}$. Since reconstruction is much easier than simulation, knowledge distillation is done by employing a feature consisting loss to minimize the distance between the hierarchical feature maps generated by the two networks.
\yingmodification{We enhance the performance of JPEG simulation by two constraints. First, we encourage the student network to minimize the difference \redmarker{in} the intermediate features generated by the last three \textit{Conv} blocks of the two networks. These features are denoted as ${\phi}_{i}, {i}\in[0,2]$, and they largely affect the final output. The constrain is commonly applied in knowledge-distillation tasks.} 
Second, we encourage both the reconstructed and simulated JPEG image, i.e., $\hat{\mathbf{I}}_\emph{jpg}'$ and $\hat{\mathbf{I}}_\emph{jpg}$, to be classified as $Q_r$ by the QF predictor. The characteristics of JPEG images can be better learned by KD-JPEG if the classification results are consistent with that of the real-world JPEG image.

The student/teacher network shares the same \yingmodification{U-shaped} Fully Convolutional Network (FCN) architecture~\cite{ronneberger2015u} with eight \textit{Conv} blocks. To \redmarker{simulate} JPEG images with \redmarker{a wide variety of} QFs, we respectively employ two individual \yingmodification{five-layered} Multi-Layer Perceptrons (MLP)
to learn the mapping functions that output modulation parameter pairs ${a}, {b}$, where \yingmodification{the deepest three layers} control the standard deviation and mean of the features.
The outputs of the leading three \textit{Conv} blocks of are \redmarker{controlled} by the MLP as follows. 
\begin{equation}
\phi_{i}={a}_{i}\cdot\emph{\redmarker{Conv\_block}}_{i}(\phi_{i-1})+{b}_{i},
\end{equation}
where $\phi_{i-1}, \phi_{i}, {a}_{i}, {b}_{i}$ represent the input and output features, the trainable mean and standard deviation at \redmarker{\textit{Conv} block} $i$, with $i\in[0,2]$. 
To implement the QF predictor, we consider that the trace for JPEG image classification lies mainly in the higher frequencies, we employ in parallel a vanilla \textit{Conv} layer, an SRM \textit{Conv} layer~\cite{srmconv} and a Bayar \textit{Conv}~\cite{bayar2018constrained} layer. \redmarker{The latter two are reported to be efficient in} depressing the main components of the input image. Thereafter, we use a typical down-sampling FCN followed by \yingmodification{three-layered} MLP. 

After training KD-JPEG, we only activate the student network and feed it with $\mathbf{X}_{\emph{tmp}}$ for JPEG simulation. Consider that $\mathbf{X}_{\emph{tmp}}$ contains more higher-band details compared to $\mathbf{I}$ because of \redmarker{information hiding}, we additionally introduce some Additive White Gaussian Noise (AWGN) on $\mathbf{I}$ and encourage the removal of higher-frequent details including AWGN during training KD-JPEG.

\noindent\textbf{Forgery detector and discriminator. }
\label{section_fogery_detector}
The forgery detector detects the tampered areas within an attacked image. When the simulation of \yingmodification{tampering} attack is skipped, we force the detector to predict a zero matrix \yingmodification{$\mathbf{O}$}. 
For the two discriminators, i.e., $\mathcal{D}_{A}$ and $\mathcal{D}_{B}$, the goals are to distinguish the generated images from the original images. High-quality immunization and recovery \redmarker{are} expected if \yingmodification{the generated images can cause misclassification of} the discriminators.

We employ the forgery detector and $\mathcal{D}_{B}$ respectively using a U-shaped network, \yingmodification{which is the same in architecture as the student/teacher network in KD-JPEG (see Fig.~\ref{fig_jpeg}) except that the MLP is not present}. That is to say, $\mathcal{D}_{B}$ conducts a pixel-wise discrimination~\cite{schonfeld2020u} on $\hat{\mathbf{I}}$ to predict which parts of the image are not recovered naturally enough. We use a simple Patch-GAN discriminator~\cite{isola2017image} to implement $\mathcal{D}_{A}$. \yingmodification{The reason we do not employ another U-shaped discriminator is that if  $\mathcal{D}_{A}$ is too strict on $\mathbf{X}$, there will not be enough space for information embedding, resulting in unstable training.}

\noindent\textbf{Implementation details. }
\yingmodification{In IINet, we apply three Haar down-sampling layers, each appended with four Double-Side Affine Coupling (DSAC) layers proposed in \cite{zhao2021invertible}, and the last conditional splitting layer is removed. The functions inside each DSAC layer can be represented by arbitrary neural networks by definition. We implement them using a five-layer residual \textit{Conv} block.}  All down-sampling and up-sampling operations are replaced with Haar down-sampling and up-sampling transformations. \yingmodification{Haar down-sampling transformation decomposes each channel of the input into four orthogonal channels, with half the width and height. Haar up-sampling transformation is the exact inverse. 
Each \textit{Conv} block contains four \textit{Conv} layers appended with a Spectral Normalization (SN)~\cite{miyato2018spectral} layer and an Exponential Linear Unit} (ELU)~\minor{\cite{ELU}} layer, and we keep the dimension of the inputs and outputs as the same. 
Exceptions are that the leading \textit{Conv} block of the U-shaped networks transforms the twelve-dimensional inputs into 32-dimensional features, and the last block transforms the 32-dimensional input features back into twelve-dimensional outputs.

\subsection{Objective Loss Function}
\label{section_loss}
The objective functions include the protection loss $\mathcal{L}_{prt}$, the \redmarker{tampering} localization loss $\mathcal{L}_{\emph{loc}}$, the recovery loss $\mathcal{L}_{\emph{rec}}$ and the nullification loss $\mathcal{L}_{\emph{null}}$. There is also the simulation loss $\mathcal{L}_{\emph{jpg}}$ exclusively for KD-JPEG. In the below equations, $\alpha,\beta,\gamma,\theta,\omega$ are the hyper-parameters.

The protection loss $\mathcal{L}_{prt}$ and the recovery loss $\mathcal{L}_{\emph{rec}}$ respectively encourage $\mathbf{X}$ and $\hat{\mathbf{I}}$ to resemble the original image $\mathbf{I}$. We use the popular $\ell_1$ loss term to measure the distance between images.
\begin{equation}
\mathcal{L}_{prt}=\lVert\mathbf{I}-\mathbf{X}\rVert_{1},
\end{equation}
\begin{equation}
\mathcal{L}_{\emph{rec}}=\lVert\mathbf{I}-\hat{\mathbf{I}}\rVert_{1}+\lVert\mathbf{E}-\hat{\mathbf{E}}\rVert_{1}.
\end{equation}

\yingmodification{
For adversarial training, the discriminators $\mathbf{D}_{A}$ and $\mathbf{D}_{B}$ need to respectively distinguish $\mathbf{X}$ and $\hat{\mathbf{I}}$ from $\mathbf{I}$. The adversarial loss $\mathcal{L}_{\emph{adv}}$ for the main pipeline is as follows.
\begin{equation}
\mathcal{L}_{\emph{adv}}=\log{(1-\mathbf{D}_{A}(\hat{\mathbf{X}}))}+\log{(1-\mathbf{D}_{B}(\hat{\mathbf{I}}))}.
\end{equation}
The loss for the two discriminators are respectively
\begin{equation}
\mathcal{L}_{\mathbf{D}_{A}}=-\frac{1}{2}(\log{\mathbf{D}_{A}(\mathbf{I})}+\log{(1-\mathbf{D}_{A}(\hat{\mathbf{X}}))}),
\end{equation}
\begin{equation}
\mathcal{L}_{\mathbf{D}_{B}}=-\frac{1}{2}(\log{\mathbf{D}_{A}(\mathbf{I})}+\log{(1-\mathbf{D}_{A}(\hat{\mathbf{I}}))}).
\end{equation}
}

The localization loss $\mathcal{L}_{\emph{loc}}$ is to \yingmodification{improve the accuracy of tampering localization.} We minimize the Binary Cross Entropy (BCE) loss between \yingmodification{the predicted mask} $\hat{\mathbf{M}}$ and \yingmodification{the ground-truth mask} $\mathbf{M}$. 
\begin{equation}
\mathcal{L}_{\emph{loc}}=-(\mathbf{M}\log{\hat{\mathbf{M}}}+(1-{\mathbf{M}})\log{(1-\hat{\mathbf{M}}})).
\end{equation}

The nullification loss $\mathcal{L}_{\emph{null}}$
is to nullify the additional output $\mathbf{Y}$ of IINet.
\begin{equation}
\label{eqn_student}
\mathcal{L}_{\emph{null}}=\lVert\mathbf{Y}-\mathbf{O}\rVert_{1}. 
\end{equation}

The total loss for \yingmodification{the main pipeline of} Imuge+ is as follows.
\begin{equation}
\mathcal{L}=\mathcal{L}_\emph{rec}+\alpha\cdot\mathcal{L}_{\emph{prt}}+\beta\cdot\mathcal{L}_{\emph{loc}}+\gamma\cdot\mathcal{L}_{\emph{null}}+\omega\cdot\mathcal{L}_{\emph{adv}}.
\end{equation}

KD-JPEG is trained ahead of the whole pipeline. Given a triple ($\mathbf{I}$,$\mathbf{I}_\emph{jpg},Q_r$), we use the Cross-Entropy (CE) loss to train the \redmarker{QF} predictor by classifying $\mathbf{I}_\emph{jpg}$.
\yingmodification{
\begin{equation}
\label{eqn_jpeg}
\mathcal{L}_{\emph{QF}}=\emph{CE}(\mathbf{Q}_{o},\mathbf{Q}_{r})=-\sum_{c=1}^{6}y_{o,c}\log(p_{o,c}),
\end{equation}
where $y$ is the binary indicator if class label $c$ is the correct classification for observation $o$. $p$ is the predicted probability observation $o$ is of class $c$. $\mathbf{Q}_{o}$ takes the \textit{argmax} of $o$ that maximizes $p$.}
For the teacher network, we employ the CE loss and the $\ell_1$ loss for reconstructing the real-world JPEG image.
\begin{equation}
\label{eqn_jpeg_teacher}
\mathcal{L}_{\emph{tea}}=\lVert\hat{\mathbf{I}}_{\emph{jpg}'}-\mathbf{I}_{\emph{jpg}}\rVert_{1}+\epsilon\cdot~\emph{CE}(\hat{\mathbf{Q}}_{r}',\mathbf{Q}_{r}).
\end{equation}
For the student network, apart from the CE loss and the $\ell_1$ loss, we additionally employ a distillation loss. The total loss for the student network is
\begin{equation}
\label{eqn_jpeg_student}
\mathcal{L}_{\emph{stu}}=\lVert\hat{\mathbf{I}}_{\emph{jpg}}-\mathbf{I}_{\emph{jpg}}\rVert_{1}+\epsilon\cdot~\emph{CE}(\hat{\mathbf{Q}}_{s},\mathbf{Q}_{r})+\sum_{i\in [\redmarker{0,2}]}\lVert{\phi}_{i}^{\emph{stu}}-{\phi}_{i}^{\emph{tea}}\rVert_{1}.
\end{equation}


\begin{figure}[!t]
	\centering
	\includegraphics[width=0.49\textwidth]{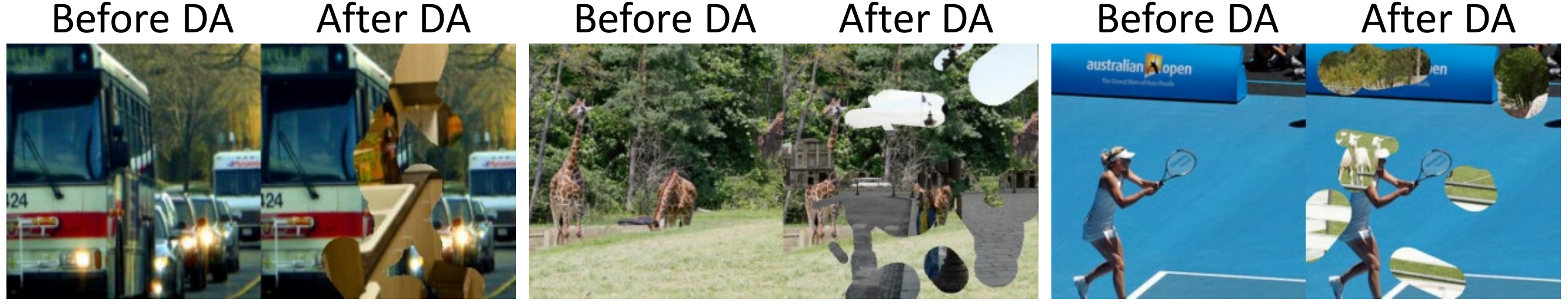}
	\caption{\textbf{Example of data augmentation by image pre-tampering.} We prevent IINet by relying on image prior by deliberately tampering some of the original images using splicing. These add-ons will be tampered again and IINet is required to recover them.}
	\label{image_data_augmentation}
\end{figure}
\subsection{Training \yingmodification{Mechanisms}}
\yingmodification{Directly training the network by minimizing $\mathcal{L}$ can hardly achieve satisfying results. The reason is mainly three-folded. First, we observe that simply varying the generated mask is not enough for effective image immunization and self-recovery, in which the network will tend to hallucinate the missing content. Secondly, we find that maintaining balanced performances under different attacks is difficult. Thirdly, a poorly-trained forgery detector will mislead the image recovery process. To address these issues, we propose the following training mechanisms.}

\noindent\textbf{Tampering-based data augmentation}.
In many cases, the randomly generated masks are not good enough to cover \yingmodification{textured areas} or the Regions of Interests (RoIs) within the images. Naturally, Deep Image Prior (DIP)~\cite{ulyanov2018deep} can be learned by networks to recover an approximate version of the original contents with semantic correctness. However, Imuge+ needs to faithfully reproduce the original image without hallucinating the results.
Therefore, how to guide the network to correctly recover the image without using DIP is a big issue.
We propose a new Data Augmentation (DA) paradigm by modifying some of the original images in the training sets using simulated splicing. Fig.~\ref{image_different_rate} shows two examples of our DA. After image immunization, we exactly tamper the add-ons during \redmarker{tampering} simulation, i.e., the mask of the first-round and second-round \redmarker{tampering} is the same. 
In other words, the introduced contents are automatically set as the \redmarker{new RoIs} of the images, and since the rest of the image shows no relation with the add-ons, IINet is forced not to use DIP for image recovery but to utilize image immunization to hide information for the recovery. We control the rate of two-round data augmentation by ${r}_{\emph{aug}}$, and empirically find that ${r}_{\emph{aug}}=15\%$ provides the best performance. The reason is that DIP can facilitate the efficient encoding of the image representation.


\noindent\textbf{Asymmetric batch size.} 
In the majority of previous deep-network-based watermarking schemes~\cite{zhu2018deep,jia2021mbrs}, the attack layer always arbitrarily and evenly performs one kind of attack on the targeted images. However, we argue that the iterative training strategy might be sub-optimal in that \yingmodification{solutions of countering different types of attacks vary. As a result,} the network upgrades \yingmodification{noisily and} unevenly \redmarker{among batches} and therefore \redmarker{can be more} in favor of \redmarker{ providing solutions for} trivial attacks. We propose to enlarge the batch size after attack simulation to balance the results among different attacks in each \redmarker{batch}. Suppose that the original batch size is $n$, we perform each of the six attacks on the $n$ images, where we get $6\cdot~n$ attacked images. Then, we concatenate these images where the batch size for the backward pass becomes $6\cdot~n$. The technique is explicitly designed to avoid disparate statistics among results produced by different image post-processing operations.

\begin{figure*}[!t]
	\centering
	\includegraphics[width=1.0\textwidth]{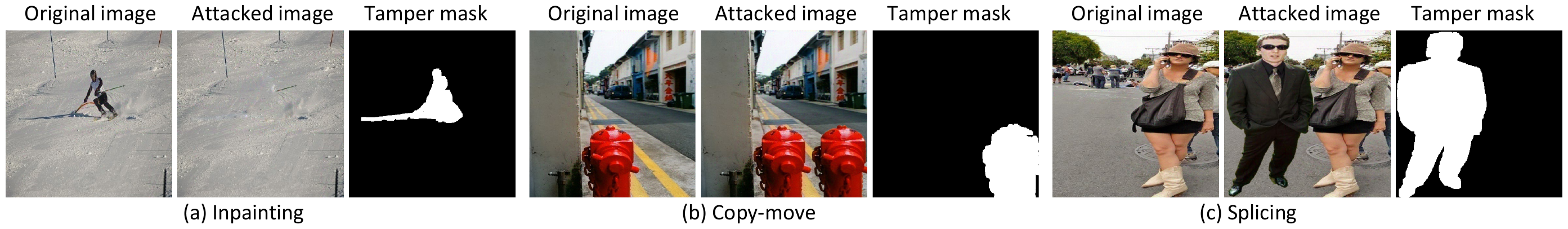}
	\caption{\textbf{Examples of the hand-crafted test set.} The images are first immunized and then manipulated by the volunteers \redmarker{who use} Adobe Photoshop and Microsoft Drawing. Malicious attacks include copy-move, inpainting and splicing. Benign attacks include typical image processings such as JPEG compression, rescaling, etc.}
	\label{img_tamper_dataset_example}
\end{figure*}
\begin{table*}[!t]
    \renewcommand{\arraystretch}{1.3}
	\caption{\textbf{Composition of \redmarker{the} real-world test dataset divided by the \redmarker{tampering} rate.} In most cases, people do not modify the images for too much, and the \redmarker{tampering} rate is generally less than 0.3. \redmarker{The settings are} consistent with that of many off-the-shelf tamper\redmarker{ing detection} datasets.}
	\label{table_dataset_distribution}
	\begin{center}
	\begin{tabular}{c|cccc|cccc|cccc}
	\hline
		\multirow{2}{*}{Attack} & \multicolumn{4}{c|}{MS-COCO} & \multicolumn{4}{c|}{ILSVRC} & \multicolumn{4}{c}{CelebA}\\
		& (0,0.1) & (0.1,0.2) & (0.2,0.3) & \textgreater0.3 & (0,0.1) & (0.1,0.2) & (0.2,0.3) & \textgreater0.3 & (0,0.1) & (0.1,0.2) & (0.2,0.3) & \textgreater0.3\\
		\hline
		Copy-move & 203 & 145 & 104 & 64 & 147 & 85 & 54 & 22 & - & - & - & -\\
		Splicing & 168 & 123 & 84 & 66 & 186 & 102 & 75 & 50 & 115 & 73 & 52 & 20\\
		Inpainting & 270 & 147 & 87 & 63 & 152 & 74 & 67 & 33 & 102 & 63 & 56 & 39 \\
		\hline
	\end{tabular}
	\end{center}
\end{table*}
\noindent\textbf{Iterative training.}.
We inherit and improve the task decoupling mechanism from~\cite{ying2021image}, since a wrongly predicted \redmarker{tampering} mask in the early training stage will unbalance the invertible function. 
We divide the training process into two stages. In the first stage, we individually train the forgery detector and the rest of the networks. 
After the generation of $\mathbf{X}_{\emph{atk}}$, the forgery detector generates the predicted \redmarker{tampering} mask $\hat{\mathbf{M}}$ and only updates itself by minimizing $\mathcal{L}_{cls}$. On the other hand, we provide IINet with the ground-truth rectified image $\hat{\mathbf{X}}_{\emph{GT}}$, where we assume a perfect \redmarker{tampering} prediction. We then update IINet by minimizing the overall loss $\mathcal{L}_{\mathcal{G}}$ with $\alpha=0$. When $\mathcal{L}_{cls}$ converges to a low level, we move on to the second stage by canceling the task decoupling, and the whole pipeline is thus trained together, where $\hat{\mathbf{X}}$ is influenced by the performance of the forgery localization. It helps IINet to adjust with imperfect \redmarker{localization} results.

\section{Experiments}
\label{section_experiment}
In this section, we conduct experiments to evaluate Imuge+. First, we clarify the experimental setup. Then we provide comprehensive experiments and analysis on \redmarker{tampering} localization and image self-recovery using Imuge+. Next, we show the performance comparison with the state-of-the-art methods and the ablation studies. Finally, we showcase three real-world applications where Imuge+ can be applied.

\subsection{Experimental Setup}
\noindent\textbf{Settings.} 
\yingmodification{
We empirically set the hyper-parameters as} $\alpha=3, \beta=\emph{1e-3}, \gamma=10, \omega=0.01$ and $\epsilon=0.1$. 
Through extensive experiments, we find that the selection of $\alpha$ and $\emph{Th}$ play important role in the ultimate network performance (see Section~\ref{section_ablation}), while the rest of the hyper-parameters have less impact.
The batch size is set as four, and we use Adam optimizer~\cite{kingma2014adam} with the default parameters. The learning rate is $1\times10^{-4}$ with the \redmarker{cosine annealing} decay. 
We binarize the prediction mask by setting the threshold $\emph{Th}$ as $0.2$. After binarization, we use image \redmarker{eroding} operation \redmarker{with kernel size $k_{\emph{ero}}=8$} for noise removal within $\hat{\mathbf{M}}$ and image dilation operation \redmarker{with kernel size $k_{\emph{dil}}=16$} to fully cover the tampered areas. 
We train Imuge+ with four distributed NVIDIA RTX 3090 \yingmodification{GPUs}. The training finishes in a week.

\noindent\textbf{Data preparation. }
Imuge+ is developed for immunizing natural images from randomized distributions, \yingmodification{being it sceneries or facial images.} Therefore, during training, the original images $\mathbf{I}$ are prepared by arbitrarily selecting around 10000 images from multiple popular datasets, namely, MS-COCO \cite{lin2014microsoft}, CelebA \cite{liu2018large}, Places \cite{zhou2017places}, UCID \cite{schaefer2003ucid} and ILSVRC \cite{johnson2016perceptual}. 
Since convolutions are generally not scale-agnostic, we train different models for some benchmark resolutions, e.g., $512\times512$, $256\times256$ and $128\times128$.
Imuge+ can be applied \redmarker{to} images with varied resolution\redmarker{s} using \redmarker{a} proper model.
The results on different resolutions are close. Therefore, the following experiments are conducted on images sized $512\times512$.
The immunized images are saved in BMP format and the attacked images are randomly saved in common formats such as JPEG, BMP or PNG. 

\noindent\textbf{Human-participated real-world \redmarker{testing}. }
Imuge+ is tested with human-participated real-world attacks where we invite several volunteers to \redmarker{manipulate} the immunized images manually using Adobe Photoshop or Microsoft 3D Drawing. 
The images for testing are from the testing part of the datasets used for training.
Note that although there are already many off-the-shelf datasets with manipulated images, such as CASIA~\cite{CASIA} and DEFACTO~\cite{DEFACTO}, Imuge+ requires image protection ahead of \redmarker{tampering} detection. Therefore, we have to first immunize intact images and request volunteers to \redmarker{manipulate} them. 
In Table~\ref{table_dataset_distribution}, we summarize the composition of the hand-crafted \redmarker{test} set. 
We group the real-world tamper\redmarker{ed} images according to the \redmarker{tampering} rate $r_T$.  
The rate of the summation of the tampered area versus the whole image is roughly $r_{\emph{sum}} \in [0.1,0.5)$, and the rate of the area of the largest \redmarker{tampering} region versus the whole image $r_{\emph{max}} \in [0.1,0.25)$, which is generally in line with that of CASIA~\cite{CASIA} and DEFACTO~\cite{DEFACTO}. 
\yingmodification{The testing set is composed of 3091 images in total.} Fig.~\ref{img_tamper_dataset_example} shows three examples.

\noindent\textbf{Evaluation metrics. }
We employ the peak signal-to-noise ratio (PSNR), the structural similarity (SSIM) \cite{wang2004image} to evaluate the image quality, and the F1 score to measure the accuracy of \redmarker{tampering} localization. 

\noindent\textbf{Benchmark. } 
We compare Imuge+ with Imuge~\cite{ying2021image} to validate the improvement in the performance of image immunzation. Additionally, there are several off-the-shelf schemes for image \redmarker{tampering} localization. We employ three state-of-the-art schemes, which detect universal image manipulations, namely, MVSS-Net~\cite{dong2021mvss}, SPAN~\cite{hu2020span} and ManTra-Net~\cite{wu2019mantra}.

\begin{table}[!t]
\renewcommand{\arraystretch}{1.2}
    \setlength{\tabcolsep}{1.5mm}
	\caption{\textbf{\yingmodification{\textbf{Comparison of computational complexity.} FLOPs: amount of floating point arithmetics. MAdd: amount of multiply-adds. MemR+W: amount of read-write memory.}}}
	\label{table_complexity}
	\begin{center}
	\begin{tabular}{c|c|c|c|c|c}
		\hline
		{Method} & {Params} & {Memory} & {MAdd} & {FLOPs} & {MemR+W} \\
        \hline
        Imuge+ & 32.0M & 5400MB & 1.53T & 0.76T & 10.92GB\\
        Imuge~\cite{ying2021image} & 13.3M & 1775MB & 0.14T & 0.28T & 3.68GB\\
        MantraNet~\cite{wu2019mantra} & 3.84M & 4706MB & 2.01T & 1.01T & 8.18GB \\
        MVSS-Net~\cite{dong2021mvss} & 142.7M & 1377MB & 0.32T & 0.16T & 3.32GB \\
		\hline
	\end{tabular}
	\end{center}
\end{table}

\begin{figure*}[!t]
	\centering
	\includegraphics[width=1.0\textwidth]{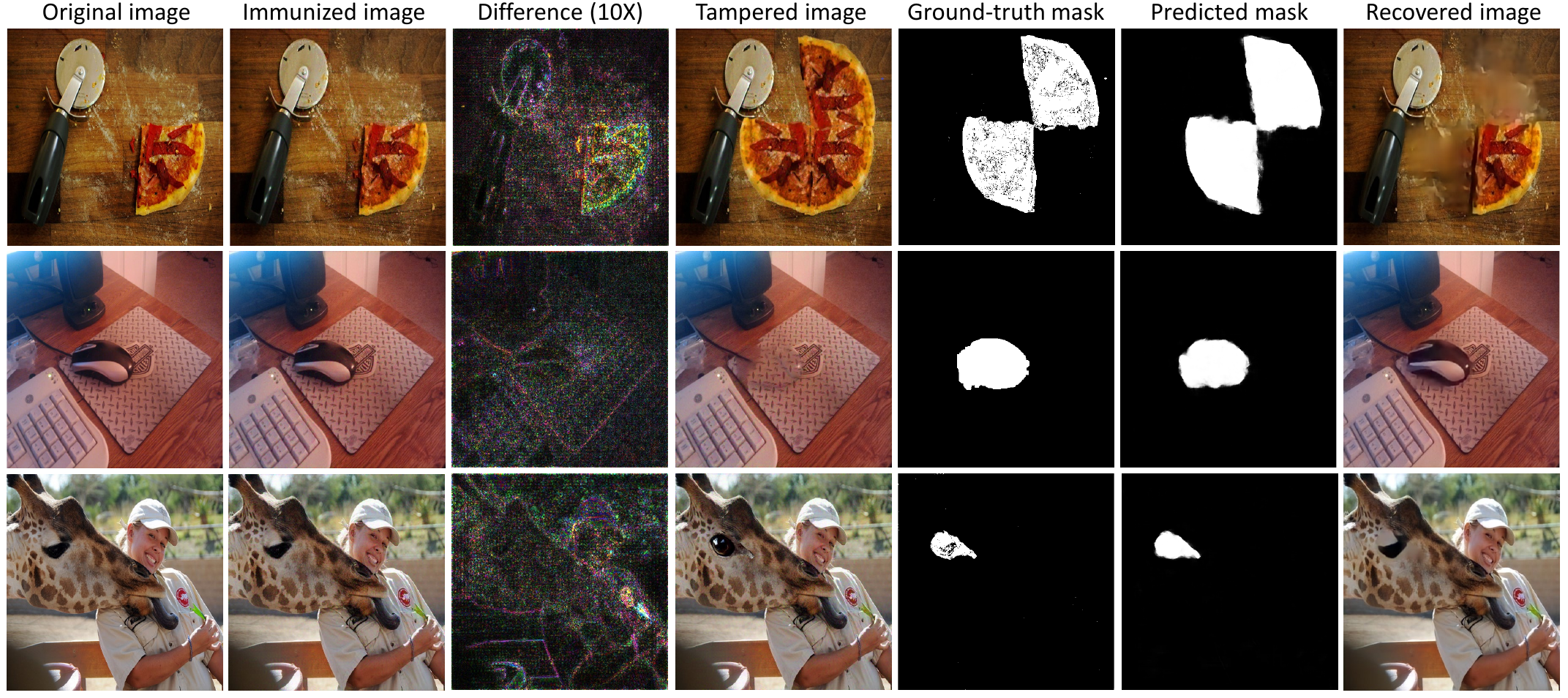}
	\caption{\textbf{Performance against combined real-world attacks.} First row: \minor{copy-move} + JPEG (QF=80). Second row: inpainting + rescaling. Third row: \minor{splicing} + Gaussian blurring. We successful conduct forgery localization and self-recovery using Imuge+.}
	\label{img_marked}
\end{figure*}

\yingmodification{
\noindent\textbf{Computational complexity. }
In Table~\ref{table_complexity}, we analyze the computational complexity of Imuge+ and compare it with Imuge, MVSS-Net and Mantra-Net. First, Imuge+ requires three times and eight times more parameters than Imuge and Mantra-Net, but much less than MVSS-Net and many other Transformer-based vision pretraining models such as Swin Transformer~\cite{liu2021swin}. Besides, the efficiency of Imuge+ is much improved compared to Imuge, where the memory cost and the amount of floating point arithmetics are comparable with Mantra-Net and MVSS-Net. Therefore, the computational complexity of Imuge+ is affordable.
}


\begin{table}[!t]
\renewcommand{\arraystretch}{1.2}
    \setlength{\tabcolsep}{1.5mm}
	\caption{\textbf{Average PSNR and SSIM between the protected images and the original images under different resolutions on MS-COCO.}}
	\label{table_different_resolution_protected}
	\begin{center}
	\begin{tabular}{c|cc|cc|cc}
		\hline
		\multirow{2}{*}{Dataset} & \multicolumn{2}{c|}{$512\times512$} & \multicolumn{2}{c|}{$256\times256$} & \multicolumn{2}{c}{$128\times128$} \\
		& PSNR & SSIM & PSNR & SSIM & PSNR & SSIM
		\\
        \hline
        COCO & 33.31& 0.945 & 33.96& 0.950 & 34.26& 0.959 \\
    
        ILSVRC & 33.48& 0.944 & 34.13& 0.951 & 34.47& 0.960 \\
    
        Places & 33.15& 0.940 & 33.77& 0.949 & 34.02& 0.953 \\
	
        UCID & 32.96& 0.937 & 33.45& 0.945 & 33.92& 0.952 \\
		\hline
	\end{tabular}
	\end{center}
\end{table}

\subsection{Real-world performances}
In Fig.~\ref{img_marked}, we randomly select three test images from MS-COCO test set. We first immunize the images and respectively invite volunteers to conduct combined attacks on them. Then Imuge+ locates the tampered areas and recovers the original image. 

\noindent\textbf{Image quality of immunized images. }
We can observe that the quality of the immunized images is satisfactory where the differences before and after image immunization are \redmarker{close to} imperceptible to human \redmarker{visual systems}. 
We have conducted more embedding experiments and the results are reported in Table~\ref{table_different_resolution_protected}, where stronger perturbations are required for datasets with textured images. For example, for MS-COCO, the average PSNR and SSIM are respectively $33.51dB$. UCID contains way more textured images, and the PSNR and SSIM slightly drop to $32.96dB$ and 0.955. 
In Table~\ref{table_different_resolution_protected}, we also conduct experiments on images with several typical resolutions. Images with smaller size enjoy a higher PSNR after image immunization, and we believe the reason is that less information is required to be self-embedded.

\begin{figure}[!t]
	\centering
	\includegraphics[width=0.49\textwidth]{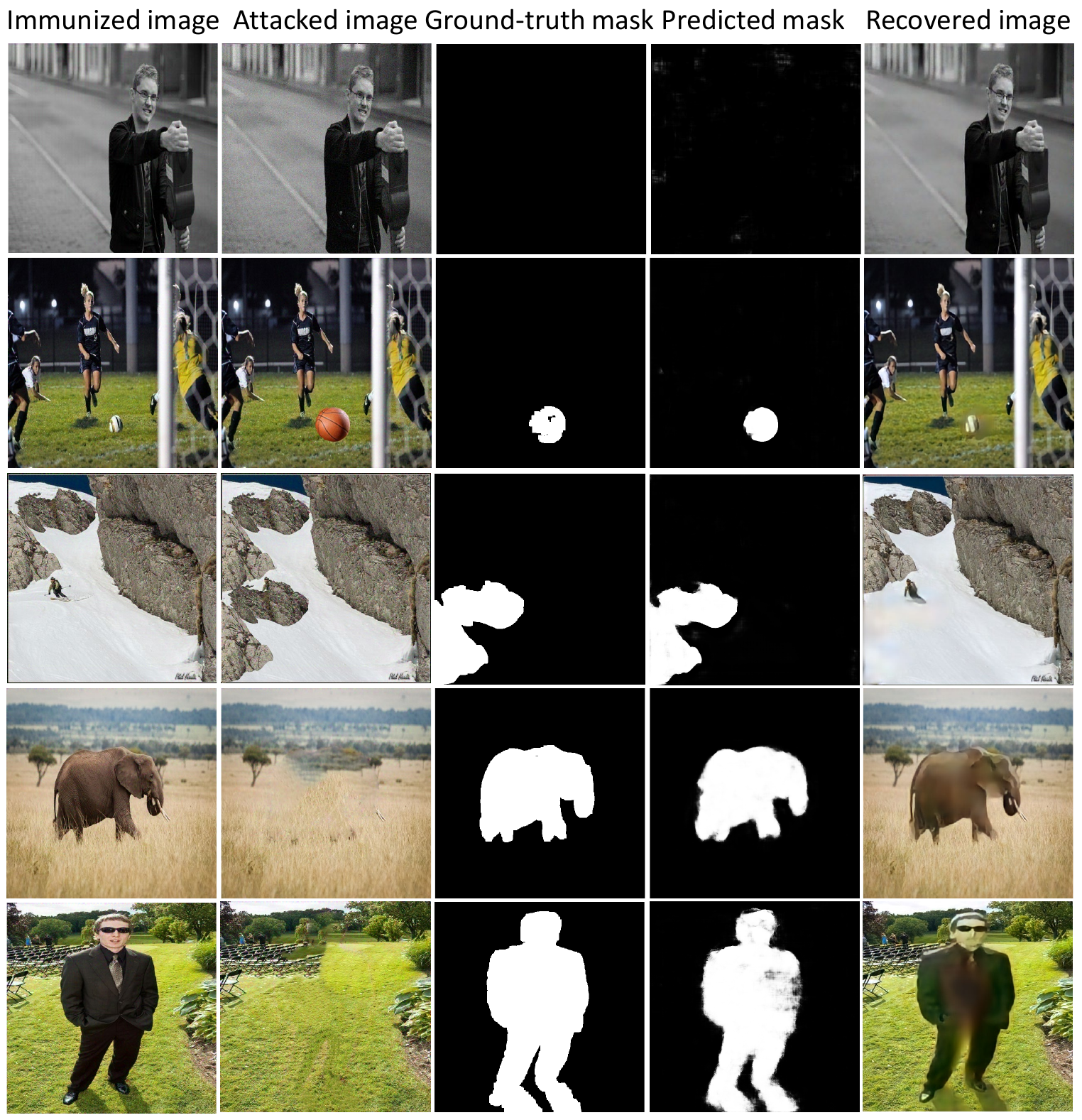}
	\caption{\textbf{Performance under various \redmarker{tampering} rate and post-processing.} The involved post-processing attacks from row one to five is respectively AWGN, Gaussian Blur, resizing, JPEG compression (QF=$80$) and JPEG compression (QF=$90$).}
	\label{image_different_rate}
\end{figure}
\begin{table*}[!t]
\renewcommand{\arraystretch}{1.2}
    \setlength{\tabcolsep}{1.5mm}
	\caption{\textbf{Performance of \redmarker{tampering} localization and image recovery tested by real-world image tampering attack.} The image \redmarker{tampering} localization and image recovery are generally satisfactory and there is no significant performance drop against image post-processing attacks.}
	\label{table_robustness}
	\begin{center}
	\begin{tabular}{c|c|c|ccc|ccc|c|cc|cc}
		\hline
		\multirow{2}{*}{Dataset} & \multirow{2}{*}{Index} & No & \multicolumn{3}{c|}{JPEG} & \multicolumn{3}{c|}{Scaling} & \multirow{2}{*}{Crop} & \multicolumn{2}{c|}{Blurring} & \multirow{2}{*}{AWGN} & \multirow{2}{*}{Drop-out}\\
		& & Attack & QF=90 & QF=70 & QF=50 & 150\% & 70\% & 50\% & & Gaussian & Median &  &  \\
        \hline
        \multirow{3}{*}{\rotatebox{90}{COCO}} 
        & F1  & 0.918 & 0.894 & 0.873 & 0.827 & 0.915 
        & 0.874 & 0.853 & 0.864 & 0.852 & 0.792 & 0.844 & 0.802 \\
        & PSNR & 29.93 & 28.87 & 28.33 & 27.82 & 29.92 & 28.45 & 27.44 & 27.61 & 28.63 & 25.65 & 27.68 &  27.13\\
        & SSIM & 0.916 & 0.905 & 0.887 & 0.873 & 0.915 & 0.864 & 0.847 & 0.854 & 0.880 & 0.787 & 0.848 &  0.847\\
		\hline
		\multirow{3}{*}{\rotatebox{90}{ILSVRC}} & F1 & 0.903 & 0.889 & 0.863 & 0.834 & 0.855 & 0.870 & 0.834 & 0.818 & 0.826 & 0.784 & 0.815 & 0.742\\
      
        & PSNR & 30.67 & 29.45 & 28.47 & 27.63 & 29.05 & 29.33 & 28.47 & 27.54 & 28.87 & 26.45 & 27.73 & 26.92\\
        & SSIM & 0.933 & 0.902 & 0.890 & 0.858 & 0.882 & 0.892 & 0.858 & 0.850 & 0.878 & 0.813 & 0.856 & 0.841 \\
		\hline
        \multirow{3}{*}{\rotatebox{90}{CelebA}} & F1 & 0.925 & 0.904 & 0.874 & 0.822 & 0.912 & 0.873 & 0.844 & 0.838 & 0.824 & 0.855 & 0.840 & 0.788\\
    
        & PSNR & 31.23 & 29.90 & 28.73 & 27.84 & 30.15 & 29.53 & 29.02 & 28.79 & 28.62 & 27.47 & 28.55 & 29.19\\
        & SSIM & 0.934 & 0.916 & 0.886 & 0.859 & 0.895 & 0.877 & 0.874 & 0.868 & 0.870 & 0.855 & 0.865 & 0.877 \\
		\hline
       
	\end{tabular}
	\end{center}
\end{table*}
\begin{table}[!t]
\renewcommand{\arraystretch}{1.2}
    \setlength{\tabcolsep}{1.5mm}
	\caption{\textbf{Average performance under different \redmarker{tampering} rate on MS-COCO.} The data before and after each slash report the performance under no attack and JPEG compression (QF=$70$).}
	\label{table_tamper_rate}
	\begin{center}
	\begin{tabular}{c|c|c|c|c}
		\hline
		$r_{T}$ & Index & Copy-move & Splicing & Inpainting\\

		\hline
		\multirow{3}{*}{[0,0.1]} & PSNR &  
		30.24 / 28.78&30.47 / 28.74&30.03 / 28.94
		 \\
				& SSIM & 
		0.917 / 0.882& 0.924 / 0.890&0.916 / 0.886
		
				\\
		& F1 &
		0.941 / 0.881&0.920 / 0.881&0.915 / 0.880
		\\
		\hline
		\multirow{3}{*}{[0.1,0.2]} & PSNR & 29.79 / 28.30&29.87 / 28.60&29.90 / 28.51
	\\
				& SSIM &
        0.910 / 0.879&0.917 / 0.883&0.915 / 0.880
		\\
		& F1 & 
        0.918 / 0.874&0.916 / 0.873&0.910 / 0.871
		\\
		\hline
		\multirow{3}{*}{[0.2,0.3]} & PSNR &
		29.57 / 28.02& 29.23 / 28.32&29.53 / 27.91
	\\
				& SSIM &
	    0.908 / 0.871&0.898 / 0.867&0.910 / 0.871
		\\
		& F1 & 
	    0.910 / 0.861&0.883 / 0.861&0.890 / 0.843
		\\
		\hline
		\multirow{3}{*}{$[0.3,0.5]$} & PSNR &
		28.97 / 27.67& 28.69 / 27.73& 28.56 / 27.36
\\
				& SSIM &
	    0.892 / 0.858 & 0.889 / 0.855 & 0.897 / 0.858
		\\
		& F1 & 
	    0.883 / 0.844&0.875 / 0.854& 0.875 / 0.822
		\\
		\hline
	\end{tabular}
	\end{center}
\end{table}
\noindent\textbf{Qualitative analysis. }
\yingmodification{As a successful localization of tampered areas is a prerequisite of successful image recovery,} Fig.~\ref{img_marked} further shows the satisfactory results of \redmarker{tampering} localization as well as image recovery. The predicted masks are close to the ground truth and the tampered contents are correctly reconstructed.
\yingmodification{In these examples, the added pizza is the result of copy-moving, and the missing mouse and eye of the giraffe is respectively removed by image inpainting and splicing.
Besides, the condition of image post-processing applied in the three examples varies, which involve the famous compression, blurring and rescaling operations. However, the diversity of the hybrid attacks still cannot prevent Imuge+ from recovering the original contents, though some details might be lost. It proves the robustness of our scheme.}

In Fig.~\ref{image_different_rate}, we showcase more examples where the tampering rate $r_{T}$ varies from zero to near 0.5. Also, these images have gone through different image post-processing methods, which are marked below the last row. When the rate is zero, we test the false alarm rate of Imuge+ where the network should identically output the provided image. The predicted mask shows that Imuge+ can evade predicting non-tampering pixels as positive even though additive AWGN is added. Besides, in the subsequent tests, the volunteers conducted the copy-move, splicing or inpainting attack respectively which result in the removal of certain non-trivial objects in the original image. From the results, we see that the tampered areas are correctly classified from the whole image plane where the borders are largely consistent. Even provided with extreme cases where $r_{T}>0.4$, Imuge+ can still recover the missing objects, though some higher-band details are lost. 

\noindent\textbf{Quantitative analysis. }
In Table~\ref{table_robustness}, we clarify the performance of Imuge+ with the presence of different image post-processing attacks. 
In Table~\ref{table_tamper_rate}, we show the average performance for different \redmarker{tampering} rate and different image post-processing attacks. Here for inpainting, the volunteers use online demo sites~\cite{yu2019free}, free tools~\cite{yu2018generative} or open-source models~\cite{nazeri2019edgeconnect}.
The tampering rate ranges from zero to $0.5$ whose distribution follows that reported in Table~\ref{table_dataset_distribution}. 
Concerning image cropping attack, Imuge+ only works within the cropped region. 

From the results, we can observe that the performance of both image \redmarker{tampering} localization and image self-recovery are successful, where the F1 score and the SSIM score are both above 0.9. Besides, Imuge+ shows strong robustness against common image post-processing behaviors in that the performances do not drop significantly despite the presence of heavy attacks such as Median blurring or AWGN addition. For example, the F1 score and the SSIM score of the worst averaged performance is near 0.8, where median blurring filters out the higher-band details of the images. In most cases, the F1 score\redmarker{s are between} 0.8 and 0.9 and the SSIM score \redmarker{between} 0.85 and 0.9. Third, it is proven by the provided results that Imuge+ can effectively work on images from different distributions, where the test images are from the most-commonly-used image datasets.
Interestingly, the performances against copy-move and splicing are slightly better than that against inpainting. The reason might be that the simulation for copy-move and splicing might be more effective than that for inpainting. Besides, the performances do not drop significantly with the increase of the \redmarker{tampering} rate.

\begin{figure}[!t]
	\centering
	\includegraphics[width=0.5\textwidth]{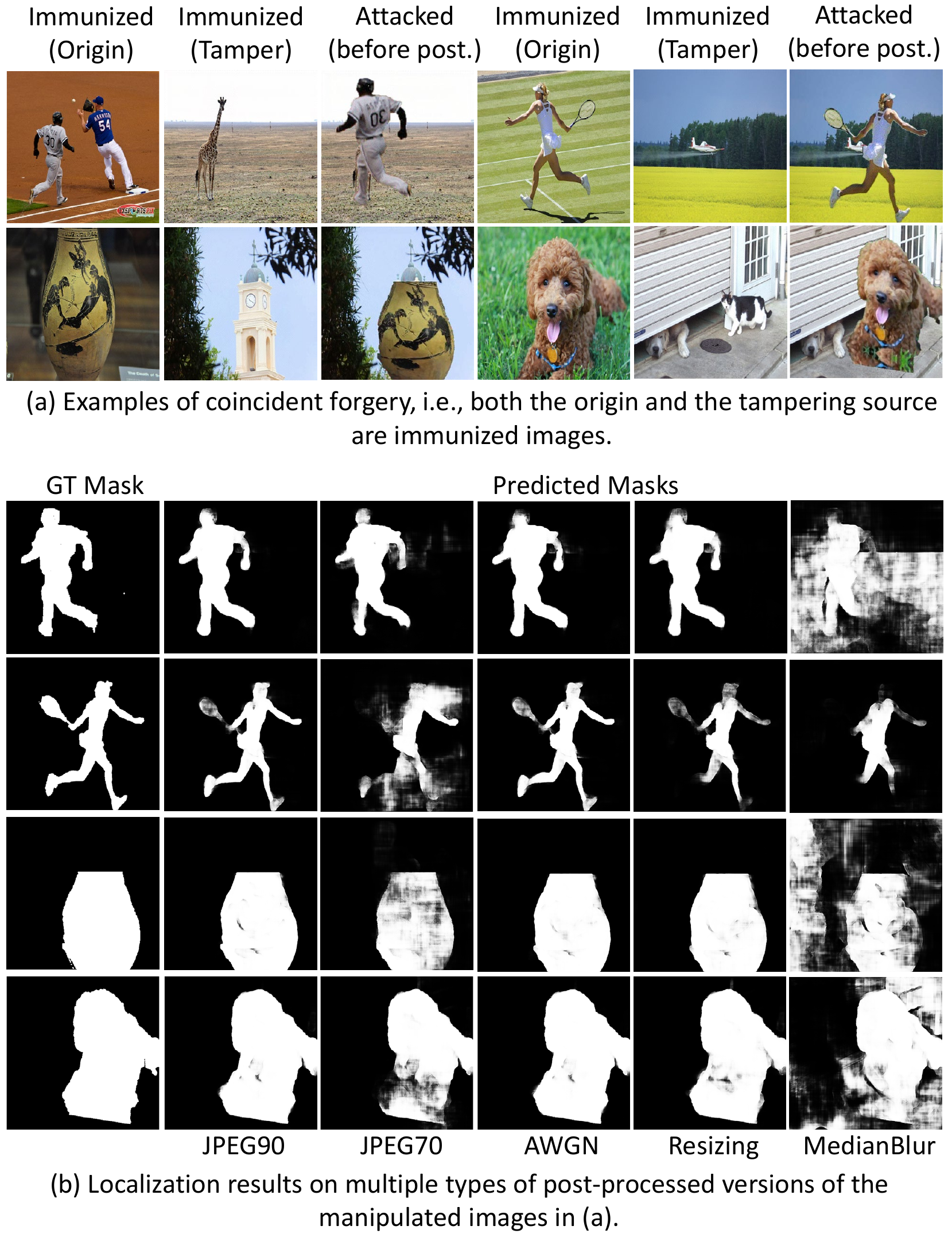}
	\caption{\minor{\textbf{Performance of Imuge+ on localization against coincident forgery.} Owing to the fact that the added items from the tampering source in most real-world application will be resized/rotated/flipped for realistic forgery, the immunization signals on them are therefore largely weakened, resulting in successful localization.}}
	\label{image_coincident}
\end{figure}
\minor{
\noindent\textbf{Coincident forgery. }
We additionally conduct experiments to test the performance of tampering localization where the tampering source used for splicing is coincidentally another immunized image.
We find that in the most common situations of image manipulation, the attacker always either resizes, flips, rotates or spatially transforms the forgery content before adding them onto the victim image, in order to produce plausible manipulation. We simulate this manipulation using two immunized images. Figure~\ref{image_coincident} provides four examples, where we observe that the coincident forgery can be detected, even though we have not included coincident forgery in the attack simulation stage.
Notice that even if the tampering rates in the last two rows exceed 50\%, the added items are not mistakenly predicted as original by the forgery detector.
The reason is that the above-mentioned distortion operations will inevitably weaken or destroy the immunized signal inside the forgery content.
}

\begin{figure}[!t]
	\centering
	\includegraphics[width=0.48\textwidth]{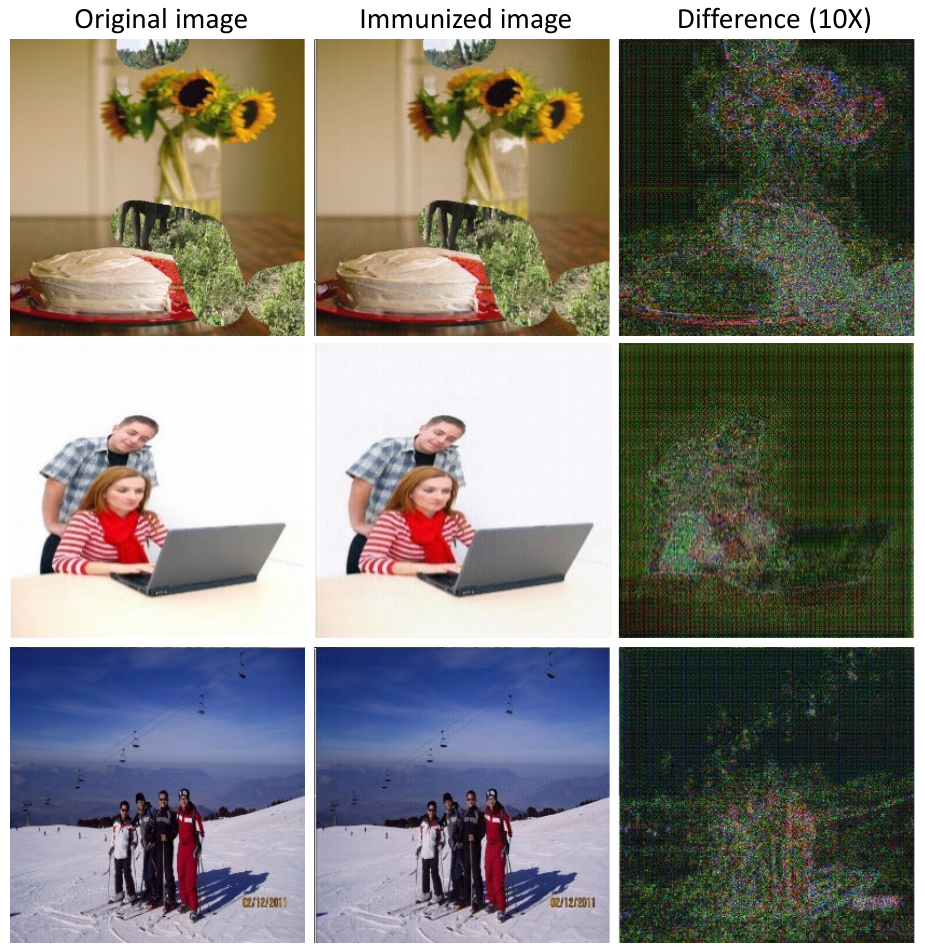}
	\caption{\yingmodification{\textbf{Analysis of the composition of immunized images.} There are two types of hidden patterns. The chess-board pattern is believed to be responsible for tampering detection, and the other is the compressed version of the original image for image self-recovery.}}
	\label{fig_immunization}
\end{figure}

\subsection{\yingmodification{What is in the immunized images?}}

\yingmodification{To analyze how Imuge+ locates the tampered areas as well as recovers the received image, we show three more examples of the immunized images in Fig.~\ref{fig_immunization} and especially have a closer observation of the augmented residual images. First, we find that in the smooth or empty areas, a checkerboard pattern is introduced by IINet, which is not so remarkable in the textured areas. We believe this chess-board pattern is embedded mainly for tampering localization, since when the immunized images are tampered, the learnt pattern is very unique and out of the distribution of natural images, which can hardly be forged or constructed by tampering attacks. Even for copy-move attacks, it will result in pattern inconsistency inside the local areas. Besides, the chess-board pattern can somehow survive JPEG compression, resizing and several kinds of blurring. Therefore, the forgery detector can detect the existence of such a pattern to determine which part of the image is forged. In contrast, state-of-the-art passive forensics schemes have to find a universally-present tampering trace, which is much harder than detecting an embedded tailored pattern. Besides, we can clearly observe that the embedded patterns near the border of the objects are different from the pattern for localization. For example, the pizza in the first example and the mouse pad in the second example are compressed into residual information and scattered around the nearby areas. We believe that on recovering the image, Imuge+ must have learned to check the surrounding area for residual information. Finally, these two kinds of patterns can harmoniously co-exist, in that we can also localize the tampered areas though the second kind of pattern is stronger in textured areas, and we believe only if the second kind of pattern is weak, will Imuge+ embed the first kind of pattern into the image.}

\begin{figure*}[!t]
	\centering
	\includegraphics[width=1.0\textwidth]{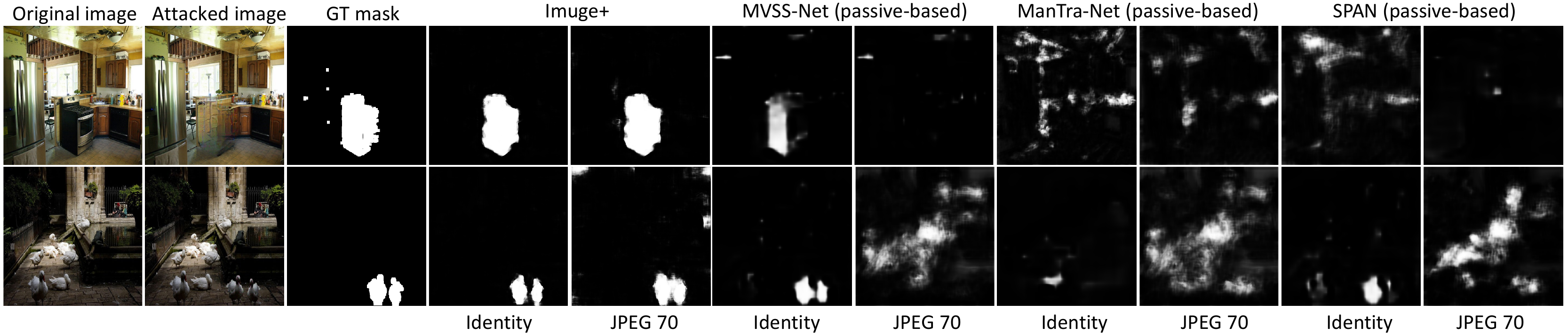}
	\caption{\textbf{Comparison of \redmarker{tampering} localization among Imuge+ and several state-of-the-art schemes.} Upper: inpainting. Lower: copy-move. Imuge+ can accurately localize the tampered areas even with the presence of post-processing attack. In contrast, many state-of-the-art schemes are reported to not have robustness.}
	\label{figure_localization_comparison}
\end{figure*}
\begin{figure}[!t]
	\centering
	\includegraphics[width=0.5\textwidth]{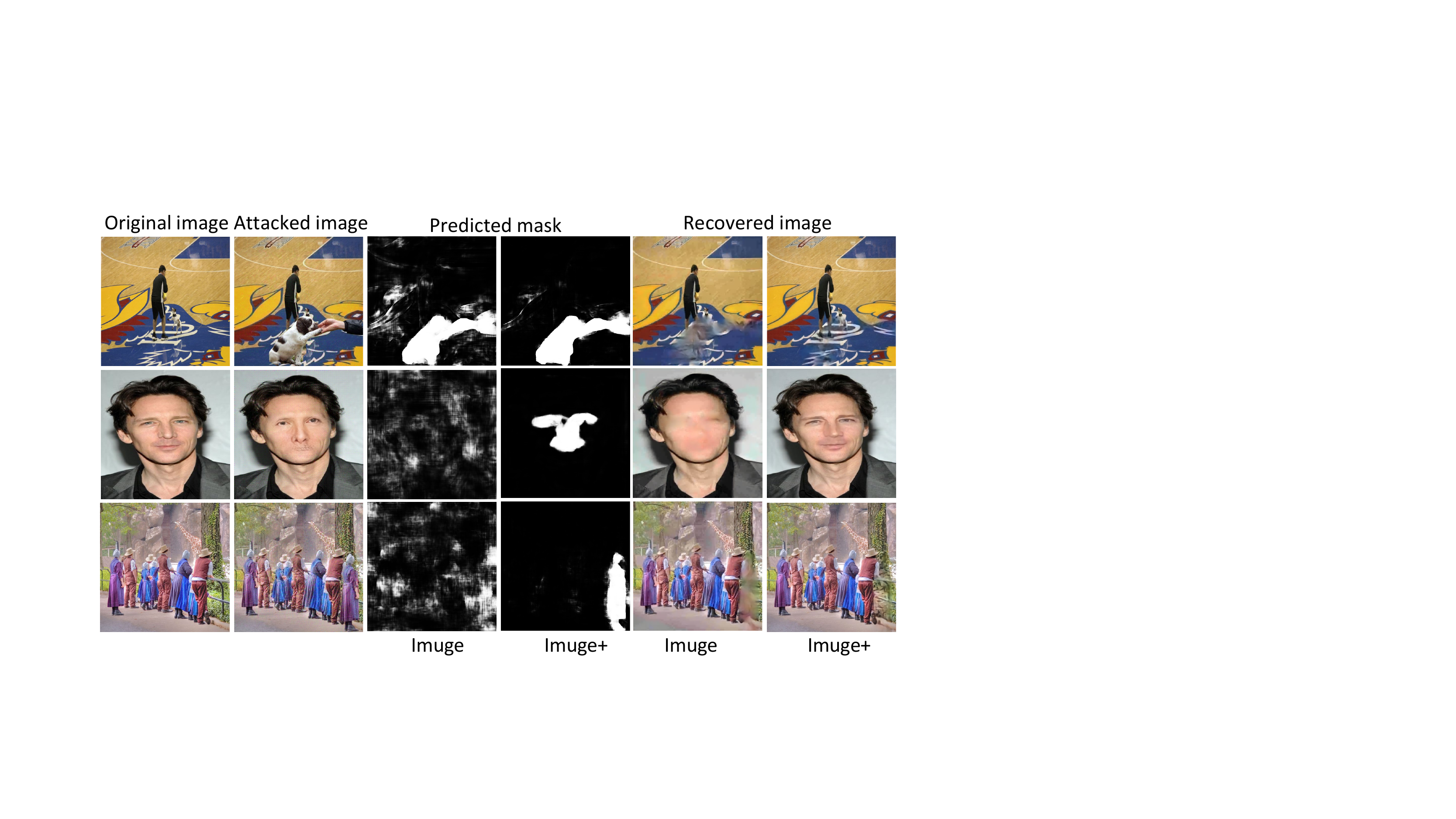}
	\caption{\textbf{Performance comparison with Imuge.} We can observe a noticeable performance boost in all three attacks. The attacked images are stored in JPEG format with QF=$70$.}
	\label{image_comparison_imuge}
\end{figure}
\begin{table}[!t]
\renewcommand{\arraystretch}{1.2}
    \setlength{\tabcolsep}{1.5mm}
	\caption{\textbf{Performance comparison with Imuge on \redmarker{tampering} localization and image recovery}. The performance of \cite{ying2021image} is in the brackets.}
	\label{table_compare_imuge}
	\begin{center}
	\begin{tabular}{c|c|c|c|c}
		\hline
		Attack & Index & Copy-move & Splicing & Inpainting\\

		\hline
		No & PSNR &  
		29.67 (27.72)& 30.13 (28,87)&29.06 (24.33)
		 \\
			
		Attack & F1 &
		0.924 (0.627)&0.903 (0.754) & 0.912 (0.431)
		\\
		\hline
		Gaussian & PSNR & 28.24 (24.49) & 28.77 (25.23) & 28.04 (23.52)
	\\
	
		Blur & F1 & 
        0.843 (0.531) & 0.866 (0.637) & 0.853 (0.333)
		\\
		\hline
		JPEG & PSNR &
		28.42 (25.37) & 28.51 (25.68)&28.03 (23.75)
	\\
		QF=70 & F1 & 
	    0.854 (0.582)& 0.881 (0.654)& 0.837 (0.295)
		\\
		\hline
	\end{tabular}
	\end{center}
\end{table}

\begin{table*}[!t]
\renewcommand{\arraystretch}{1.2}
	\setlength{\tabcolsep}{1.5mm}
	\caption{\textbf{F1 score comparison for \redmarker{tampering} detection among our scheme and the state-of-the-art methods.} Our scheme ranks first in all the tests and leads by a large margin in robustness. }
	\label{table_localization}
    \begin{center}
    \begin{tabular}{c|cccc|cccc|cccc}
    \hline
    		\multirow{2}{*}{Method} & \multicolumn{4}{c|}{ Inpainting} & \multicolumn{4}{c|}{Splicing} & \multicolumn{4}{c}{Copy-move}\\
    		& NoAttack & JPEG & Blurring & Scaling & NoAttack & JPEG & Blurring & Scaling & NoAttack & JPEG & Blurring & Scaling\\
    		\hline
    		Imuge~\cite{ying2021image} & 0.372 & 0.253 & 0.219 & 0.327 & 0.754 & 0.712 & 0.679 & 0.747 & 0.642 & 0.463 & 0.453 & 0.566\\
    		Mantra-Net~\cite{wu2019mantra}  & 0.584 & 0.540 & 0.517 & 0.559 & 0.678 & 0.563 & 0.504 & 0.539 & 0.553 & 0.477 & 0.423 & 0.511\\
    		MVSS-Net~\cite{dong2021mvss}& 0.673 & 0.487 & 0.445 & 0.570 & 0.787 & 0.659 & 0.535 & 0.672 & 0.701 & 0.627 & 0.551 & 0.675\\
    		SPAN~\cite{hu2020span}& 0.701 & 0.624  & 0.599 & 0.683 & 0.732 & 0.657 & 0.574 & 0.665 & 0.685 & 0.614 & 0.565 & 0.648\\
    		Imuge+ & \textbf{0.917} & \textbf{0.872}  & \textbf{0.854} & \textbf{0.885} & \textbf{0.923} & \textbf{0.878} & \textbf{0.849} & \textbf{0.873} & \textbf{0.903} & \textbf{0.844} & \textbf{0.829} & \textbf{0.837}\\
    		\hline
    	\end{tabular}
    	\end{center}
\end{table*}
\subsection{Comparison}

\noindent\textbf{Content recovery within tampered areas. }
We compare Imuge+ with \cite{ying2021image} \yingmodification{to verify the performance boost brought by our enhanced network design and novel training mechanisms.} \yingmodification{The averaged PSNRs of the two methods between $\mathbf{I}$ and $\mathbf{X}$ are kept close for a fair comparison.} Besides, in order to compare the overall quality of image recovery, assume that the \redmarker{tampering} localization is correct in measuring the performance of image recovery. 

In Fig.~\redmarker{\ref{image_comparison_imuge}}, we show \redmarker{three groups of experimental} comparison between Imuge and Imuge+, where we perform \redmarker{splicing} attack in the first row, \redmarker{inpainting} attack in the second and \yingmodification{copy-move attack in the third row}. 
In Table~\ref{table_compare_imuge}, we provide the \redmarker{averaged} results \redmarker{of the comparison}.
Notice that the face of the man in the \yingmodification{second} example is completely removed, but Imuge+ can successfully \redmarker{recover the original face}. Besides, Imuge is not trained against image inpainting attack and therefore it performs poorly in the \yingmodification{second} example.
From the averaged results, where we see that the performance boost is noticeable. First, Imuge+ is more accurate in localizing the tampers. Second, Imuge+ can preserve more detail of the missing objects. \yingmodification{Third, Imuge+ is stronger in overall robustness against various kinds of image post-processing attacks.}

\noindent\textbf{Image \redmarker{tampering} localization. }
\yingmodification{Passive image forensics schemes do not rely on hiding additional information, but they usually find tiny traces that are vulnerable to traditional image post-processing, so the performances can deteriorate in real-world OSN applications.}
\yingmodification{Though tampering localization in Imuge+ requires an additional and mandatory procedure of information hiding, our scheme aims at replacing original images which are prone to a variety of hybrid tampering and post-processing attacks with their immunized versions. Therefore, by restricting the magnitude of embedded perturbation, we wish the immunized images to be viewed as \textit{original images} in the future.}

\yingmodification{To better evaluate the accuracy of tampering localization of our scheme,} here we \yingmodification{briefly} compare the performance with those of the state-of-the-art \yingmodification{passive image tampering localization schemes}.
Note that for \cite{dong2021mvss,kwon2021cat,wu2019mantra}, we conduct fair comparisons by performing the same attacks on the original images, \yingmodification{not the immunized images}.
Fig.~\redmarker{\ref{figure_localization_comparison}} shows two comparison results where we respectively involve inpainting and copy-move attacks.

According to the results, Imuge+ provides the leading performance which is robust to the JPEG compression attack with QF=$70$. In contrast, passive methods encounter a significant performance drop.
In Table~\ref{table_localization}, we further show the averaged accuracy of the above-mentioned schemes. Generally, the F1 scores of Imuge+ are above 0.8, which shows high resilience against JPEG compression, blurring and scaling. The overall performance is significantly improved compared to \cite{ying2021image} which performs only fair against copy-move attack and poorly against inpainting. The testing results of \cite{dong2021mvss,kwon2021cat,wu2019mantra} are consistent with those recorded in these papers. The average F1 score of these methods under no attack is within [0.55,0.8], \redmarker{and the performances} under \yingmodification{image post-processing} attacks drop by 0.1 to 0.2.


\begin{figure}[!t]
	\centering
	\includegraphics[width=0.49\textwidth]{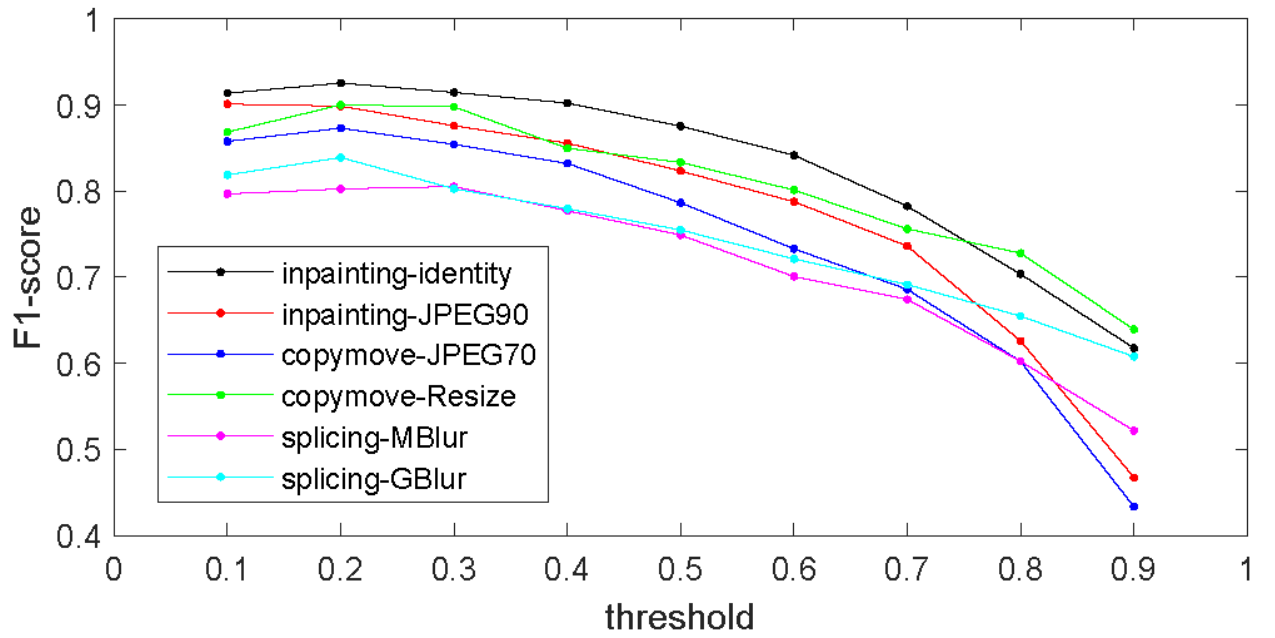}
	\caption{\textbf{Performance curves of F1 score versus threshold.} Generally, F1 reaches the highest peak when the threshold is $0.2$.}
	\label{image_F1}
\end{figure}
\begin{figure}[!t]
	\centering
	\includegraphics[width=0.49\textwidth]{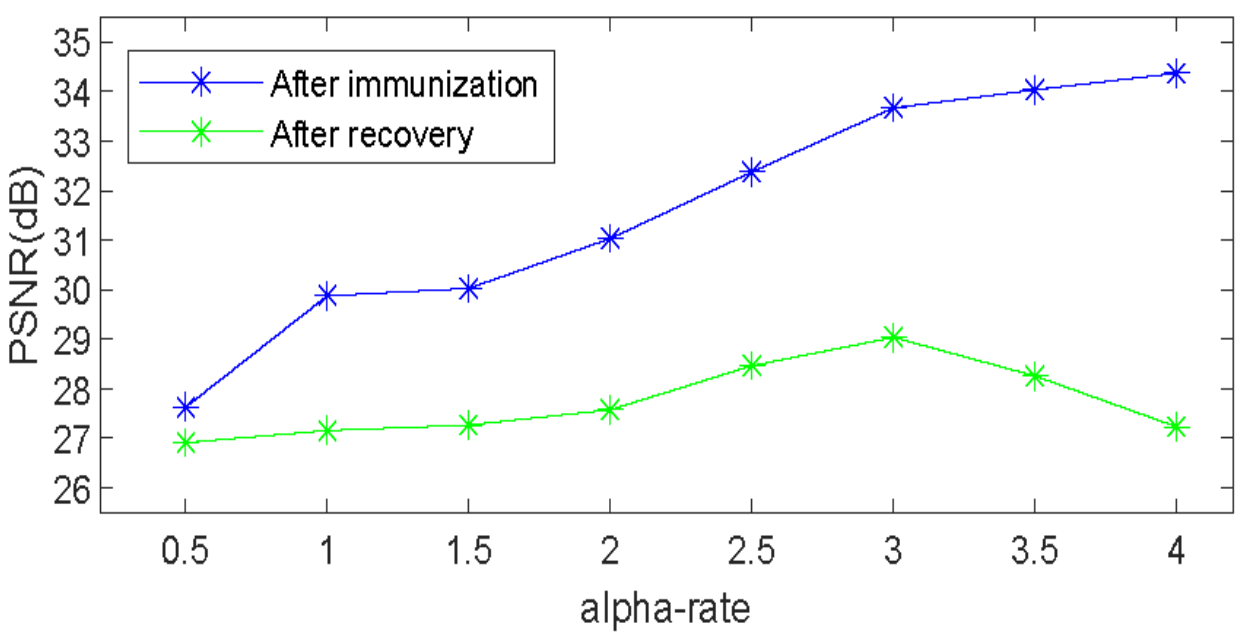}
	\caption{\textbf{Performance curves of PSNR versus $\alpha$}. We observe that the trade-off between imperceptibility and network effectiveness can be best balanced when $\alpha$ equals to $3$.}
	\label{image_thresh}
\end{figure}
\subsection{Ablation Study}
\label{section_ablation}
\begin{figure*}[!t]
	\centering
	\includegraphics[width=1.0\textwidth]{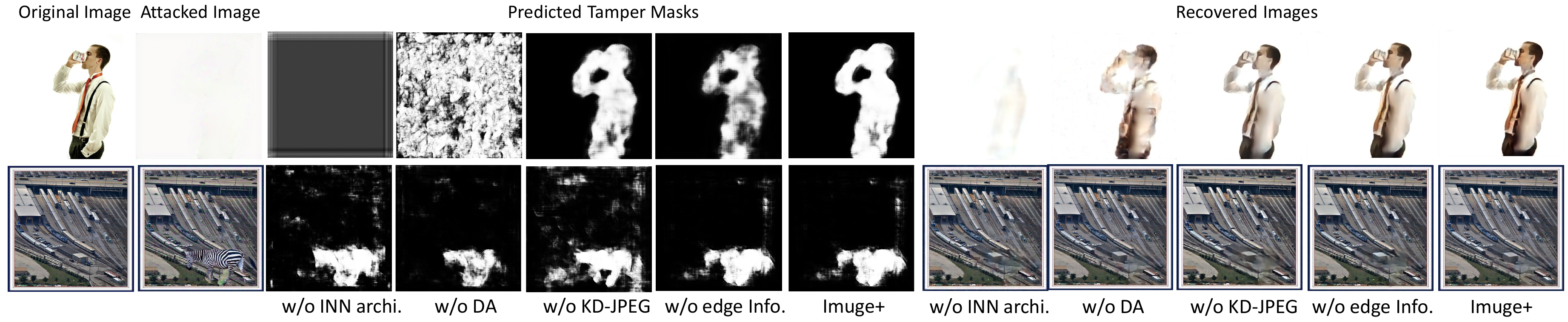}
	\caption{\textbf{Ablation study of Imuge+.} In each group, the performed attack is the same, \yingmodification{with JPEG QF=80}. Among all of the setups, Imuge+ with full setup performs the best in both \redmarker{tampering} localization and image recovery.}
	\label{figure_ablation}
\end{figure*}

We discuss the selection of the hyper-parameters and the threshold for binarizing the prediction mask, respectively in Fig.~\ref{image_F1} and Fig.~\ref{image_thresh}. 
Besides, we explore the influence of the key components in Imuge+ by evaluating the performance of the model with varied and partial setups. 
In each test, we remove a single component and train the models from scratch till they converge.
Fig.~\ref{figure_ablation} visualizes two groups of performance comparison under hybrid attacks. 
We also summarize the average result on partial setups in Table~\ref{table_ablation} by performing the same image-processing attacks and the same tampering attacks in each test.

\begin{figure*}[!t]
	\centering
	\includegraphics[width=1.0\textwidth]{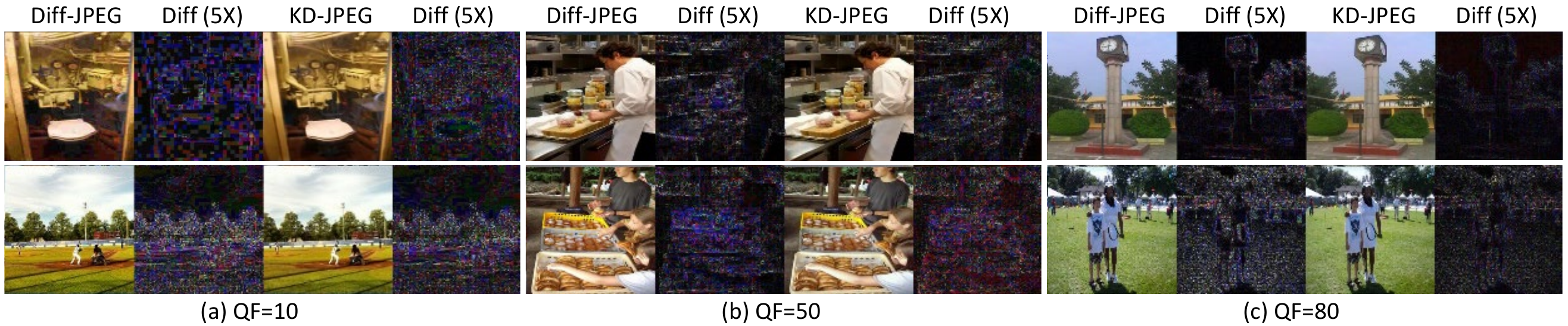}
	\caption{\textbf{Comparison on the fidelity of the simulated JPEG images.} \yingmodification{Owing to the flexibility of the generative network,} KD-JPEG in average contains less distortion compared to Diff-JPEG~\cite{shin2017jpeg}.}
	\label{figure_kdjpeg_result}
\end{figure*}
\begin{table}[!t]
\renewcommand{\arraystretch}{1.2}
	\caption{\textbf{Ablation study of Imuge+ using varied partial settings.} The tests are done under JPEG attack (QF=$70$) and $r_{T}\in[0.1,0.4]$. \minor{'IP': image protection (forward pass of IINet). 'E': including edge as additional input and supervision. 'KD': using KD-JPEG as JPEG simulator. 'TDA': tampering-based DA. 'IT': iterative training. 'AB': asymmetric batch size. $\Delta^1$: using Diff-JPEG~\cite{shin2017jpeg} as JPEG simulator. $\Delta^2$: using two separate U-Net~\cite{ronneberger2015u} to replace IINet. '-': failed to train steadily.}}
    \begin{center}
    \begin{tabular}{ccc|ccc|c|c|c}
    \hline 
    		\multicolumn{6}{c}{Network components} & \multicolumn{3}{c}{Index}\\
    		\hline
    		IP&E&KD&TDA&IT&AB&F1&PSNR&SSIM\\
    		\hline
    		&\checkmark&\checkmark&\checkmark&\checkmark&\checkmark& 0.435 & 18.13  & 0.585\\
    		\hline
    		$\Delta^1$&\checkmark&\checkmark&\checkmark&\checkmark&\checkmark& 0.933 & 21.54 & 0.713 \\
    		\hline
    		\checkmark& &\checkmark&\checkmark&\checkmark&\checkmark & 0.854 & 24.57  & 0.784\\
    		\hline
    		\checkmark&\checkmark&$\Delta^2$&\checkmark&\checkmark&\checkmark & 0.962 & 23.78  & 0.755\\
    		\hline
    		\checkmark&\checkmark&\checkmark& &\checkmark&\checkmark & 0.435 & 24.63  & 0.744 \\
    		\hline
    		\checkmark&\checkmark&\checkmark&\checkmark& &\checkmark & - & -  & -\\
    		\hline
    		\checkmark&\checkmark&\checkmark&\checkmark&\checkmark& & 0.913 & 25.59  & 0.810\\
    		\hline
    		\checkmark&\checkmark&\checkmark&\checkmark&\checkmark&\checkmark & $\textbf{0.965}$ & $\textbf{26.41}$  & $\textbf{0.822}$ \\
    		\hline
    	\end{tabular}
    	\end{center}
    	\label{table_ablation}
\end{table}

\noindent\textbf{Selection of the critical hyper-parameters. }
In Fig.~\ref{image_F1}, we study the selection of proper threshold for predicted mask binarization $\emph{Th}$ and rate of data augmentation ${r}_{\emph{aug}}$. From the curves, we find that when $\emph{Th}$ is $0.2$, we can generally acquire the best performance. Besides, we also empirically find that the performance can be further enhanced when ${r}_{\emph{aug}}$ is set as $15\%$.
In Fig.~\ref{image_thresh}, we also study the trade-off between the imperceptibility of image protection and the effectiveness of image recovery. Generally, when the PSNR \redmarker{surpasses} $33dB$, there \redmarker{will be little} easily-noticeable artifacts caused by the immunization. According to both the human visual system and the figure, $\alpha$ is set as three for the best equilibrium.
To begin with, we verify that directly conducting forgery detection and image recovery on the attacked \textit{unimmunized} images leads to sub-optimal results.

\noindent\textbf{Influence of the JPEG simulator.}
Fig~\ref{figure_kdjpeg_result} shows the comparison between KD-JPEG and Diff-JPEG~\cite{shin2017jpeg} for JPEG simulation. In each group, we respectively compare the absolute difference between the generated JPEG images and the real-world JPEG images. Our generative method gives closer results to the real-world JPEG in that the differences are smaller and the characteristics of JPEG compression are better studied. The checkerboard artifact of JPEG compression can also be found in the generated JPEG images.
On average, the PSNR between the generated JPEG images by KD-JPEG and the real-world JPEG images is $29.64dB$ when QF is $10$, and that of Diff-JPEG is $28.03dB$. 
Moreover, the PSNR of KD-JPEG is $31.17dB$ when QF is $50$ and $34.42dB$ when QF is $90$. In comparison, Diff-JPEG shows less flexibility in the simulation where the PSNR is $30.22dB$ when QF is $50$ and $33.50dB$ when QF is $90$.
Besides, the QF classification accuracy of images generated by KD-JPEG is $95.47\%$, while that of the differentiable method Diff-JPEG~\cite{shin2017jpeg} $72.48\%$. 
It indicates that more feature representations of the JPEG images can be learned by KD-JPEG.
Fig.~\ref{figure_ablation} shows that applying Diff-JPEG leads to more blurry results. 

\noindent\textbf{Influence of the INN-based architecture.}
\yingmodification{The benefit of introducing INN-based architectures for invertible function modeling has been well studied by many researches~\cite{kingma2018glow,dinh2014nice}.} These networks learn deterministic and invertible distribution mapping, where the forward and back propagation operations are in the same network. A typical alternative is to model image protection and image recovery independently using the well-known ``Encoder-Decoder" architecture. From the results, we observe that INN-based Imuge+ provides much better performance. The reason is that INN has much fewer hyper-parameters and therefore \yingmodification{the training process is stable compared to GAN training.}

\noindent\textbf{Influence of the data augmentation. }
We find that without using our \redmarker{tampering}-based data augmentation, the network during training still tends to use DIP for image recovery. The reason is that the randomly generated masks may catch the plain-text areas where the contents inside are trivial compared to the surroundings. 
In the first example of Fig.~\ref{figure_ablation}, \yingmodification{we see that without the novel data augmentation, Imuge+ can still somehow recover the removed person, suggesting that even if without disabling semantic hints, Imuge+ still reconstructs the images using the hidden information.}
However, we see a much lower quality and fidelity compared to the full implementation. As we introduce irrelevant information and force the network to tamper and recover them, Imuge+ cannot always rely on DIP and tries to embed more essential information for reliable recovery.

\noindent\textbf{Influence of other components. }
\yingmodification{Previous work~\cite{nazeri2019edgeconnect} has proven that the recovery of intermediate representation, like edges and the gray-scaled version, can boost the performance of image reconstruction.}
Compared to \cite{ying2021image}, Imuge+ additionally embeds and recovers the edge information to enforce that the recovered results are semantically and trustfully correct. According to the ablation studies, without introducing the edge supervision, \yingmodification{Imuge+ tends to produce more blurry recovered images}, while there is no noticeable effect on the \redmarker{tampering} localization.
Besides, we find that without using iterative training, we cannot steadily train Imuge+ where poorly predicted tampering masks heavily disrupts the image recovery stage. 
Also, performing the asymmetric-batch-size technique can help noticeable promoting the overall performance.

\begin{figure}[!t]
	\centering
	\includegraphics[width=0.49\textwidth]{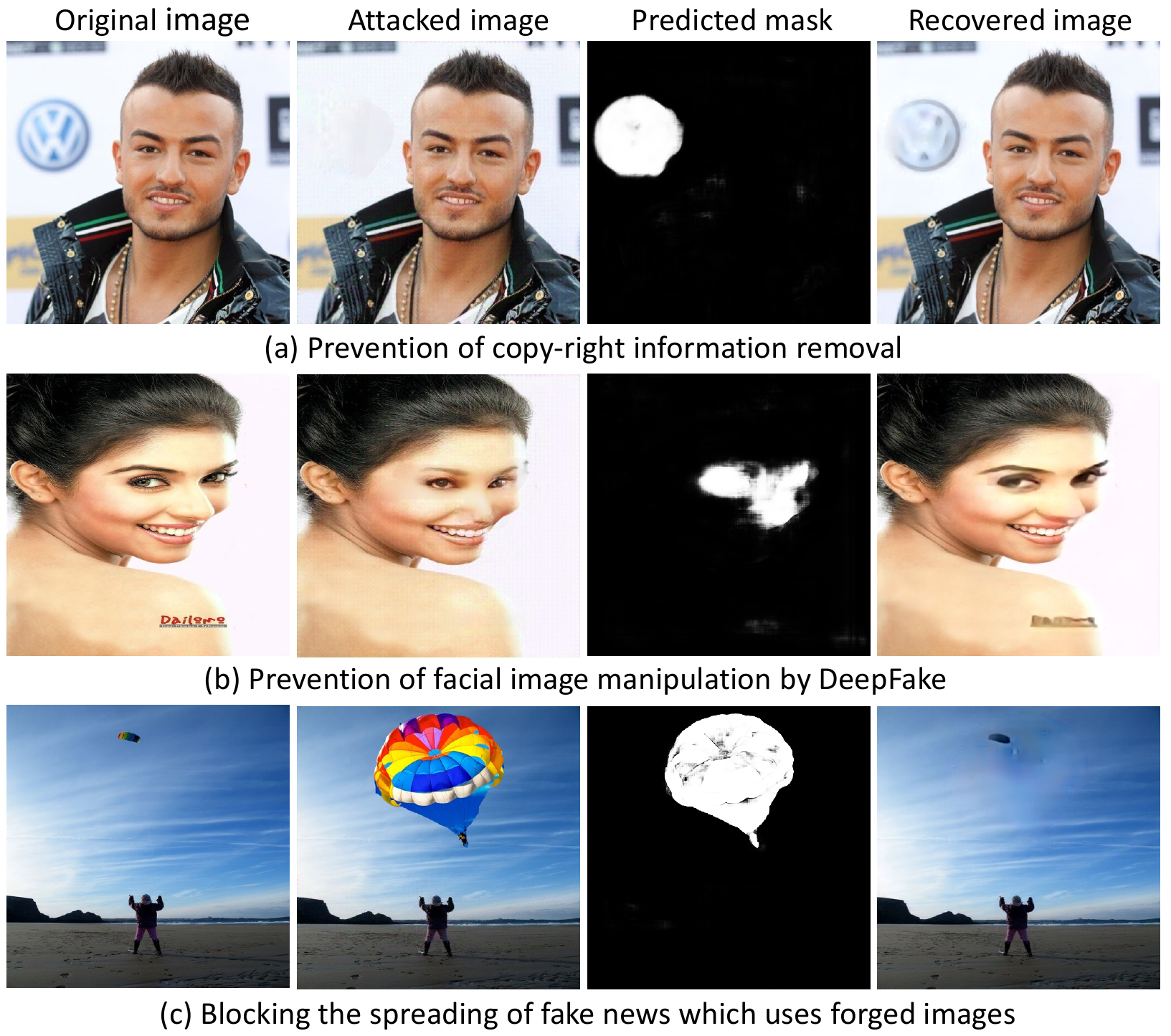}
	\caption{\textbf{Applications of Imuge+ against fake images.} The proposed method is effective in both localizing the tampered area and recovering the original content.}
	\label{image_application}
\end{figure}

\subsection{Applications}
\yingmodification{Image immunization is tested on thousands of hand-crafted tampered images. It is designed for real-world automatic image tampering localization and self-recovery. }
In Fig.~\ref{image_application}, we exemplify three applications of Imuge+ for different purposes in image forensics. 

\noindent\textbf{Prevention of copy-right information removal. }
\redmarker{Several} image tampering attacks are based on the removal of critical information, such as signatures, logos, etc. Such behaviors are a serious violation of copy-right protection and image fidelity. Using image immunization, Imuge+ can benefit the prevention of their removal. For example, in the first row of the figure, the logo of Volkswagen was removed by the attacker and later identified as well as recovered in high quality by Imuge+. 

\noindent\textbf{Prevention of facial image manipulation by DeepFake. }
The second example is to combat image manipulation by DeepFake. Facial images are also prone to modifications that illegally shift the identity of the person involved. Such tampering can be easily accomplished by GAN-based technologies and result in severe security issues. In the second row, Imuge+ predicts that the face was forged by image inpainting technique and successfully reconstructs the identifiable original face. In that sense, Imuge+ also blocks the spreading of fake news which uses forged images as supplementary multimedia.

\noindent\textbf{Reversible image editing. }
Normally, after we edit an image, a copy of the original version has to be stored if we wish to revert the unwanted editing later. However, storing every version as a backup after each tiny modification might be \redmarker{expensive}. In the third example, the data owner removes the added parachute, and uses Imuge+ to revert the editing without having stored the original image. In the recovered image, the missing tiny parachute is also recovered. Therefore, the users are no longer required to store the original version of an image after modification. We leave it a future work to develop a repeatable image immunization that can revert each modification individually with only one immunized version stored.

\section{Conclusions \yingmodification{ and Future Works}}
In this paper, we present a novel generative scheme called Imuge+, which is an image tamper resilient generative scheme for image self-recovery. We transform the original images into immunized images, where the tamper attacks can be accurately localized and the original content within the tampered areas can be recovered. To boost the performance, we form the invertible function for image immunization and employ an INN architecture for implementation. Besides, we propose a novel JPEG simulator as well as an enhanced attack layer for greater resilience against common image post-processing attacks. We conduct comprehensive experiments on several popular datasets and invite several volunteers to manipulate the immunized images, and the results prove the effectiveness of Imuge+ in both tamper localization and content recovery against splicing, copy-move and inpainting attacks.

\yingmodification{There are still some remaining issues to be addressed in future works. 
First, the imperceptibility of the information embedded required by image immunization can be further improved. 
We find that we cannot immunize an immunized image, otherwise the corresponding generated image will be disastrous in quality.
Second, though we have made steady improvements in the overall quality of the recovered images, the blurry issue still exists in many cases, e.g., the attacked images are heavily compressed, or a single tampered region is too big resulting in the center area not recoverable.
Third, IMUGE+ is still largely at black-bos level, and therefore, we wish to conduct more theoretical analysis, especially on the upper bound for image immunization against common types of OSN attack.
We hope that in the near future, the above-mentioned issues can be addressed well, and Imuge+ can be integrated into cameras, so that image immunization can be introduced into the image signal processing pipeline, thereby changing the current situation where digital images can be freely edited.}




\section*{Acknowledgments}
This work is supported by National Natural Science Foundation of China under Grant U20B2051, U1936214. We are grateful to the anonymous reviewers for their constructive and insightful suggestions on improving this paper.


\bibliographystyle{IEEEtran}
\bibliography{reference}

\begin{thebibliography}{100}
\providecommand{\url}[1]{#1}
\csname url@samestyle\endcsname
\providecommand{\newblock}{\relax}
\providecommand{\bibinfo}[2]{#2}
\providecommand{\BIBentrySTDinterwordspacing}{\spaceskip=0pt\relax}
\providecommand{\BIBentryALTinterwordstretchfactor}{4}
\providecommand{\BIBentryALTinterwordspacing}{\spaceskip=\fontdimen2\font plus
\BIBentryALTinterwordstretchfactor\fontdimen3\font minus
  \fontdimen4\font\relax}
\providecommand{\BIBforeignlanguage}[2]{{%
\expandafter\ifx\csname l@#1\endcsname\relax
\typeout{** WARNING: IEEEtran.bst: No hyphenation pattern has been}%
\typeout{** loaded for the language `#1'. Using the pattern for}%
\typeout{** the default language instead.}%
\else
\language=\csname l@#1\endcsname
\fi
#2}}
\providecommand{\BIBdecl}{\relax}
\BIBdecl

\bibitem{ijcv2011}
I.~Yerushalmy and H.~Hel-Or, ``Digital image forgery detection based on lens
  and sensor aberration,'' \emph{International Journal of Computer Vision
  (IJCV)}, vol.~92, no.~1, pp. 71--91, 2011.

\bibitem{liu2021swin}
Z.~Liu, Y.~Lin, Y.~Cao, H.~Hu, Y.~Wei, Z.~Zhang, S.~Lin, and B.~Guo, ``Swin
  transformer: Hierarchical vision transformer using shifted windows,'' in
  \emph{Proceedings of the IEEE/CVF International Conference on Computer
  Vision}, 2021, pp. 10\,012--10\,022.

\bibitem{suvorov2022resolution}
R.~Suvorov, E.~Logacheva, A.~Mashikhin, A.~Remizova, A.~Ashukha, A.~Silvestrov,
  N.~Kong, H.~Goka, K.~Park, and V.~Lempitsky, ``Resolution-robust large mask
  inpainting with fourier convolutions,'' in \emph{Proceedings of the IEEE/CVF
  Winter Conference on Applications of Computer Vision}, 2022, pp. 2149--2159.

\bibitem{ELU}
D.-A. Clevert, T.~Unterthiner, and S.~Hochreiter, ``Fast and accurate deep
  network learning by exponential linear units (elus),'' \emph{arXiv preprint
  arXiv:1511.07289}, 2015.

\bibitem{yu2018generative}
J.~Yu, Z.~Lin, J.~Yang, X.~Shen, X.~Lu, and T.~S. Huang, ``Generative image
  inpainting with contextual attention,'' in \emph{Proceedings of the IEEE/CVF
  Conference on Computer Vision and Pattern Recognition (CVPR)}, 2018, pp.
  5505--5514.

\bibitem{wang2019musical}
N.~Wang, J.~Li, L.~Zhang, and B.~Du, ``Musical: Multi-scale image contextual
  attention learning for inpainting.'' in \emph{IJCAI}, 2019, pp. 3748--3754.

\bibitem{nazeri2019edgeconnect}
K.~Nazeri, E.~Ng, T.~Joseph, F.~Qureshi, and M.~Ebrahimi, ``Edgeconnect:
  Structure guided image inpainting using edge prediction,'' in
  \emph{Proceedings of the IEEE/CVF International Conference on Computer Vision
  Workshops (ICCVW)}, 2019, pp. 0--0.

\bibitem{srmconv}
P.~Zhou, X.~Han, V.~I. Morariu, and L.~S. Davis, ``Learning rich features for
  image manipulation detection,'' in \emph{Proceedings of the IEEE/CVF
  Conference on Computer Vision and Pattern Recognition (CVPR)}, 2018, pp.
  1053--1061.

\bibitem{cozzolino2015splicebuster}
------, ``Learning rich features for image manipulation detection,'' in
  \emph{Proceedings of the IEEE/CVF Conference on Computer Vision and Pattern
  Recognition (CVPR)}, 2018, pp. 1053--1061.

\bibitem{MATLAB:2018}
MATLAB, \emph{9.7.0.1190202 (R2019b)}.\hskip 1em plus 0.5em minus 0.4em\relax
  Natick, Massachusetts: The MathWorks Inc., 2018.

\bibitem{yu2019free}
J.~Yu, Z.~Lin, J.~Yang, X.~Shen, X.~Lu, and T.~S. Huang, ``Free-form image
  inpainting with gated convolution,'' in \emph{Proceedings of the IEEE/CVF
  International Conference on Computer Vision (ICCV)}, 2019, pp. 4471--4480.

\bibitem{wang2021image}
T.~Wang, H.~Ouyang, and Q.~Chen, ``Image inpainting with external-internal
  learning and monochromic bottleneck,'' in \emph{Proceedings of the IEEE/CVF
  Conference on Computer Vision and Pattern Recognition (CVPR)}, 2021, pp.
  5120--5129.

\bibitem{zhuang2021image}
P.~Zhuang, H.~Li, S.~Tan, B.~Li, and J.~Huang, ``Image tampering localization
  using a dense fully convolutional network,'' \emph{IEEE Transactions on
  Information Forensics and Security (TIFS)}, vol.~16, pp. 2986--2999, 2021.

\bibitem{li2019localization}
H.~Li and J.~Huang, ``Localization of deep inpainting using high-pass fully
  convolutional network,'' in \emph{Proceedings of the IEEE/CVF International
  Conference on Computer Vision (ICCV)}, 2019, pp. 8301--8310.

\bibitem{jo2021rethinking}
Y.~Jo, S.~Y. Chun, and J.~Choi, ``Rethinking deep image prior for denoising,''
  in \emph{Proceedings of the IEEE/CVF International Conference on Computer
  Vision (ICCV)}, 2021, pp. 5087--5096.

\bibitem{guan2019mfc}
H.~Guan, M.~Kozak, E.~Robertson, Y.~Lee, A.~N. Yates, A.~Delgado, D.~Zhou,
  T.~Kheyrkhah, J.~Smith, and J.~Fiscus, ``Mfc datasets: Large-scale benchmark
  datasets for media forensic challenge evaluation,'' in \emph{2019 IEEE Winter
  Applications of Computer Vision Workshops (WACVW)}.\hskip 1em plus 0.5em
  minus 0.4em\relax IEEE, 2019, pp. 63--72.

\bibitem{hsu2006detecting}
Y.-F. Hsu and S.-F. Chang, ``Detecting image splicing using geometry invariants
  and camera characteristics consistency,'' in \emph{2006 IEEE International
  Conference on Multimedia and Expo}.\hskip 1em plus 0.5em minus 0.4em\relax
  IEEE, 2006, pp. 549--552.

\bibitem{liao2020robust}
X.~Liao, K.~Li, X.~Zhu, and K.~R. Liu, ``Robust detection of image operator
  chain with two-stream convolutional neural network,'' \emph{IEEE Journal of
  Selected Topics in Signal Processing}, vol.~14, no.~5, pp. 955--968, 2020.

\bibitem{yang2020constrained}
C.~Yang, H.~Li, F.~Lin, B.~Jiang, and H.~Zhao, ``Constrained r-cnn: A general
  image manipulation detection model,'' in \emph{2020 IEEE International
  Conference on Multimedia and Expo (ICME)}.\hskip 1em plus 0.5em minus
  0.4em\relax IEEE, 2020, pp. 1--6.

\bibitem{zhu2018deep}
X.~Zhu, Y.~Qian, X.~Zhao, B.~Sun, and Y.~Sun, ``A deep learning approach to
  patch-based image inpainting forensics,'' \emph{Signal Processing: Image
  Communication}, vol.~67, pp. 90--99, 2018.

\bibitem{li2018fast}
Y.~Li and J.~Zhou, ``Fast and effective image copy-move forgery detection via
  hierarchical feature point matching,'' \emph{IEEE Transactions on Information
  Forensics and Security (TIFS)}, vol.~14, no.~5, pp. 1307--1322, 2018.

\bibitem{salloum2018image}
R.~Salloum, Y.~Ren, and C.-C.~J. Kuo, ``Image splicing localization using a
  multi-task fully convolutional network (mfcn),'' \emph{Journal of Visual
  Communication and Image Representation}, vol.~51, pp. 201--209, 2018.

\bibitem{kwon2021cat}
M.-J. Kwon, I.-J. Yu, S.-H. Nam, and H.-K. Lee, ``Cat-net: Compression artifact
  tracing network for detection and localization of image splicing,'' in
  \emph{Proceedings of the IEEE/CVF Winter Conference on Applications of
  Computer Vision (WACV)}, 2021, pp. 375--384.

\bibitem{wu2017deep}
Y.~Wu, W.~Abd-Almageed, and P.~Natarajan, ``Deep matching and validation
  network: An end-to-end solution to constrained image splicing localization
  and detection,'' in \emph{Proceedings of the 25th ACM international
  conference on Multimedia}, 2017, pp. 1480--1502.

\bibitem{rao2016deep}
Y.~Rao and J.~Ni, ``A deep learning approach to detection of splicing and
  copy-move forgeries in images,'' in \emph{2016 IEEE International Workshop on
  Information Forensics and Security (WIFS)}.\hskip 1em plus 0.5em minus
  0.4em\relax IEEE, 2016, pp. 1--6.

\bibitem{cozzolino2015efficient}
D.~Cozzolino, G.~Poggi, and L.~Verdoliva, ``Efficient dense-field copy--move
  forgery detection,'' \emph{IEEE Transactions on Information Forensics and
  Security (TIFS)}, vol.~10, no.~11, pp. 2284--2297, 2015.

\bibitem{van2019reversible}
T.~F. van~der Ouderaa and D.~E. Worrall, ``Reversible gans for memory-efficient
  image-to-image translation,'' in \emph{Proceedings of the IEEE/CVF Conference
  on Computer Vision and Pattern Recognition (CVPR)}, 2019, pp. 4720--4728.

\bibitem{wang2020modeling}
Y.~Wang, M.~Xiao, C.~Liu, S.~Zheng, and T.-Y. Liu, ``Modeling lost information
  in lossy image compression,'' \emph{arXiv preprint arXiv:2006.11999}, 2020.

\bibitem{lugmayr2020srflow}
A.~Lugmayr, M.~Danelljan, L.~Van~Gool, and R.~Timofte, ``Srflow: Learning the
  super-resolution space with normalizing flow,'' in \emph{European Conference
  on Computer Vision (ECCV)}.\hskip 1em plus 0.5em minus 0.4em\relax Springer,
  2020, pp. 715--732.

\bibitem{ardizzone2019guided}
L.~Ardizzone, C.~L{\"u}th, J.~Kruse, C.~Rother, and U.~K{\"o}the, ``Guided
  image generation with conditional invertible neural networks,'' \emph{arXiv
  preprint arXiv:1907.02392}, 2019.

\bibitem{boehm2014stegexpose}
B.~Boehm, ``Stegexpose-a tool for detecting lsb steganography,'' \emph{arXiv
  preprint arXiv:1410.6656}, 2014.

\bibitem{jacobsen2018revnet}
J.-H. Jacobsen, A.~Smeulders, and E.~Oyallon, ``i-revnet: Deep invertible
  networks,'' \emph{arXiv preprint arXiv:1802.07088}, 2018.

\bibitem{behrmann2019invertible}
J.~Behrmann, W.~Grathwohl, R.~T. Chen, D.~Duvenaud, and J.-H. Jacobsen,
  ``Invertible residual networks,'' in \emph{International Conference on
  Machine Learning (ICML)}, 2019, pp. 573--582.

\bibitem{shin2017jpeg}
R.~Shin and D.~Song, ``Jpeg-resistant adversarial images,'' in \emph{NIPS 2017
  Workshop on Machine Learning and Computer Security}, vol.~1, 2017.

\bibitem{hu2020span}
X.~Hu, Z.~Zhang, Z.~Jiang, S.~Chaudhuri, Z.~Yang, and R.~Nevatia, ``Span:
  Spatial pyramid attention network for image manipulation localization,'' in
  \emph{European Conference on Computer Vision (ECCV)}.\hskip 1em plus 0.5em
  minus 0.4em\relax Springer, 2020, pp. 312--328.

\bibitem{bengio2013estimating}
Y.~Bengio, N.~L{\'e}onard, and A.~Courville, ``Estimating or propagating
  gradients through stochastic neurons for conditional computation,''
  \emph{arXiv preprint arXiv:1308.3432}, 2013.

\bibitem{zhang2021towards}
C.~Zhang, A.~Karjauv, P.~Benz, and I.~S. Kweon, ``Towards robust deep hiding
  under non-differentiable distortions for practical blind watermarking,'' in
  \emph{Proceedings of the 29th ACM International Conference on Multimedia},
  2021, pp. 5158--5166.

\bibitem{clevert2015fast}
D.-A. Clevert, T.~Unterthiner, and S.~Hochreiter, ``Fast and accurate deep
  network learning by exponential linear units (elus),'' \emph{arXiv preprint
  arXiv:1511.07289}, 2015.

\bibitem{bayar2018constrained}
B.~Bayar and M.~C. Stamm, ``Constrained convolutional neural networks: A new
  approach towards general purpose image manipulation detection,'' \emph{IEEE
  Transactions on Information Forensics and Security (TIFS)}, vol.~13, no.~11,
  pp. 2691--2706, 2018.

\bibitem{tao2018towards}
J.~Tao, S.~Li, X.~Zhang, and Z.~Wang, ``Towards robust image steganography,''
  \emph{IEEE Transactions on Circuits and Systems for Video Technology
  (TCSVT)}, vol.~29, no.~2, pp. 594--600, 2018.

\bibitem{chen2017deeplab}
L.-C. Chen, G.~Papandreou, I.~Kokkinos, K.~Murphy, and A.~L. Yuille, ``Deeplab:
  Semantic image segmentation with deep convolutional nets, atrous convolution,
  and fully connected crfs,'' \emph{IEEE Transactions on Pattern Analysis and
  Machine Intelligence (TPAMI)}, vol.~40, no.~4, pp. 834--848, 2017.

\bibitem{long2015fully}
J.~Long, E.~Shelhamer, and T.~Darrell, ``Fully convolutional networks for
  semantic segmentation,'' in \emph{Proceedings of the IEEE/CVF Conference on
  Computer Vision and Pattern Recognition (CVPR)}, 2015, pp. 3431--3440.

\bibitem{sedighi2015content}
V.~Sedighi, R.~Cogranne, and J.~Fridrich, ``Content-adaptive steganography by
  minimizing statistical detectability,'' \emph{IEEE Transactions on
  Information Forensics and Security (TIFS)}, vol.~11, no.~2, pp. 221--234,
  2015.

\bibitem{dong2021mvss}
X.~Chen, C.~Dong, J.~Ji, J.~Cao, and X.~Li, ``Image manipulation detection by
  multi-view multi-scale supervision,'' in \emph{Proceedings of the IEEE/CVF
  International Conference on Computer Vision (ICCV)}, 2021, pp.
  14\,185--14\,193.

\bibitem{chen2018defining}
K.~Chen, H.~Zhou, W.~Zhou, W.~Zhang, and N.~Yu, ``Defining cost functions for
  adaptive jpeg steganography at the microscale,'' \emph{IEEE Transactions on
  Information Forensics and Security (TIFS)}, vol.~14, no.~4, pp. 1052--1066,
  2018.

\bibitem{fang2018screen}
H.~Fang, W.~Zhang, H.~Zhou, H.~Cui, and N.~Yu, ``Screen-shooting resilient
  watermarking,'' \emph{IEEE Transactions on Information Forensics and Security
  (TIFS)}, vol.~14, no.~6, pp. 1403--1418, 2018.

\bibitem{agarwal2019survey}
N.~Agarwal, A.~K. Singh, and P.~K. Singh, ``Survey of robust and imperceptible
  watermarking,'' \emph{Multimedia Tools and Applications}, vol.~78, no.~7, pp.
  8603--8633, 2019.

\bibitem{asikuzzaman2017overview}
M.~Asikuzzaman and M.~R. Pickering, ``An overview of digital video
  watermarking,'' \emph{IEEE Transactions on Circuits and Systems for Video
  Technology (TCSVT)}, vol.~28, no.~9, pp. 2131--2153, 2017.

\bibitem{shi2016reversible}
Y.-Q. Shi, X.~Li, X.~Zhang, H.-T. Wu, and B.~Ma, ``Reversible data hiding:
  advances in the past two decades,'' \emph{IEEE access}, vol.~4, pp.
  3210--3237, 2016.

\bibitem{baluja2017hiding}
S.~Baluja, ``Hiding images in plain sight: Deep steganography,'' \emph{Advances
  in Neural Information Processing Systems (NIPS)}, vol.~30, pp. 2069--2079,
  2017.

\bibitem{baluja2019hiding}
------, ``Hiding images within images,'' \emph{IEEE Transactions on Pattern
  Analysis and Machine Intelligence (TPAMI)}, 2019.

\bibitem{duan2019reversible}
X.~Duan, K.~Jia, B.~Li, D.~Guo, E.~Zhang, and C.~Qin, ``Reversible image
  steganography scheme based on a u-net structure,'' \emph{IEEE Access},
  vol.~7, pp. 9314--9323, 2019.

\bibitem{rahim2018end}
R.~Rahim, S.~Nadeem \emph{et~al.}, ``End-to-end trained cnn encoder-decoder
  networks for image steganography,'' in \emph{Proceedings of the European
  Conference on Computer Vision (ECCV)}, 2018, pp. 0--0.

\bibitem{mun2019finding}
S.-M. Mun, S.-H. Nam, H.~Jang, D.~Kim, and H.-K. Lee, ``Finding robust domain
  from attacks: A learning framework for blind watermarking,''
  \emph{Neurocomputing}, vol. 337, pp. 191--202, 2019.

\bibitem{kandi2017exploring}
H.~Kandi, D.~Mishra, and S.~R.~S. Gorthi, ``Exploring the learning capabilities
  of convolutional neural networks for robust image watermarking,''
  \emph{Computers \& Security}, vol.~65, pp. 247--268, 2017.

\bibitem{zhu2018hidden}
J.~Zhu, R.~Kaplan, J.~Johnson, and L.~Fei-Fei, ``Hidden: Hiding data with deep
  networks,'' in \emph{Proceedings of the European conference on computer
  vision (ECCV)}, 2018, pp. 657--672.

\bibitem{luo2020distortion}
X.~Luo, R.~Zhan, H.~Chang, F.~Yang, and P.~Milanfar, ``Distortion agnostic deep
  watermarking,'' in \emph{Proceedings of the IEEE/CVF Conference on Computer
  Vision and Pattern Recognition (CVPR)}, 2020, pp. 13\,548--13\,557.

\bibitem{bi2020d}
X.~Bi, Y.~Liu, B.~Xiao, W.~Li, C.-M. Pun, G.~Wang, and X.~Gao, ``D-unet: A
  dual-encoder u-net for image splicing forgery detection and localization,''
  \emph{arXiv preprint arXiv:2012.01821}, 2020.

\bibitem{ronneberger2015u}
O.~Ronneberger, P.~Fischer, and T.~Brox, ``U-net: Convolutional networks for
  biomedical image segmentation,'' in \emph{International Conference on Medical
  Image Computing and Computer-Assisted Intervention (MICCAI)}.\hskip 1em plus
  0.5em minus 0.4em\relax Springer, 2015, pp. 234--241.

\bibitem{szegedy2016rethinking}
C.~Szegedy, V.~Vanhoucke, S.~Ioffe, J.~Shlens, and Z.~Wojna, ``Rethinking the
  inception architecture for computer vision,'' in \emph{Proceedings of the
  IEEE/CVF Conference on Computer Vision and Pattern Recognition (CVPR)}, 2016,
  pp. 2818--2826.

\bibitem{isola2017image}
P.~Isola, J.-Y. Zhu, T.~Zhou, and A.~A. Efros, ``Image-to-image translation
  with conditional adversarial networks,'' in \emph{Proceedings of the IEEE/CVF
  Conference on Computer Vision and Pattern Recognition (CVPR)}, 2017, pp.
  1125--1134.

\bibitem{johnson2016perceptual}
J.~Johnson, A.~Alahi, and L.~Fei-Fei, ``Perceptual losses for real-time style
  transfer and super-resolution,'' in \emph{European Conference on Computer
  Vision (ECCV)}.\hskip 1em plus 0.5em minus 0.4em\relax Springer, 2016, pp.
  694--711.

\bibitem{rudin1992nonlinear}
L.~I. Rudin, S.~Osher, and E.~Fatemi, ``Nonlinear total variation based noise
  removal algorithms,'' \emph{Physica D: nonlinear phenomena}, vol.~60, no.
  1-4, pp. 259--268, 1992.

\bibitem{lin2014microsoft}
T.-Y. Lin, M.~Maire, S.~Belongie, J.~Hays, P.~Perona, D.~Ramanan,
  P.~Doll{\'a}r, and C.~L. Zitnick, ``Microsoft coco: Common objects in
  context,'' in \emph{European Conference on Computer Vision (ECCV)}.\hskip 1em
  plus 0.5em minus 0.4em\relax Springer, 2014, pp. 740--755.

\bibitem{deng2009imagenet}
J.~Deng, W.~Dong, R.~Socher, L.-J. Li, K.~Li, and L.~Fei-Fei, ``Imagenet: A
  large-scale hierarchical image database,'' in \emph{2009 IEEE/CVF Conference
  on Computer Vision and Pattern Recognition (CVPR)}.\hskip 1em plus 0.5em
  minus 0.4em\relax Ieee, 2009, pp. 248--255.

\bibitem{wang2004image}
Z.~Wang, A.~C. Bovik, H.~R. Sheikh, and E.~P. Simoncelli, ``Image quality
  assessment: from error visibility to structural similarity,'' \emph{IEEE
  Transactions on Image Processing (TIP)}, vol.~13, no.~4, pp. 600--612, 2004.

\bibitem{yu2020attention}
C.~Yu, ``Attention based data hiding with generative adversarial networks,'' in
  \emph{Proceedings of the AAAI Conference on Artificial Intelligence},
  vol.~34, no.~01, 2020, pp. 1120--1128.

\bibitem{mao2017least}
X.~Mao, Q.~Li, H.~Xie, R.~Y. Lau, Z.~Wang, and S.~Paul~Smolley, ``Least squares
  generative adversarial networks,'' in \emph{Proceedings of the IEEE
  International Conference on Computer Vision (ICCV)}, 2017, pp. 2794--2802.

\bibitem{ioffe2015batch}
S.~Ioffe and C.~Szegedy, ``Batch normalization: Accelerating deep network
  training by reducing internal covariate shift,'' in \emph{International
  Conference on Machine Learning (ICML)}, 2015, pp. 448--456.

\bibitem{carlini2017towards}
N.~Carlini and D.~Wagner, ``Towards evaluating the robustness of neural
  networks,'' in \emph{2017 ieee symposium on security and privacy (sp)}.\hskip
  1em plus 0.5em minus 0.4em\relax IEEE, 2017, pp. 39--57.

\bibitem{zhang2017mixup}
H.~Zhang, M.~Cisse, Y.~N. Dauphin, and D.~Lopez-Paz, ``mixup: Beyond empirical
  risk minimization,'' \emph{arXiv preprint arXiv:1710.09412}, 2017.

\bibitem{khachaturov2021markpainting}
D.~Khachaturov, I.~Shumailov, Y.~Zhao, N.~Papernot, and R.~Anderson,
  ``Markpainting: Adversarial machine learning meets inpainting,'' in
  \emph{International Conference on Machine Learning (ICML)}, 2021, pp.
  5409--5419.

\bibitem{yin2018deep}
M.~Yin, Y.~Zhang, X.~Li, and S.~Wang, ``When deep fool meets deep prior:
  Adversarial attack on super-resolution network,'' in \emph{Proceedings of the
  26th ACM international conference on Multimedia}, 2018, pp. 1930--1938.

\bibitem{ying2021image}
Q.~Ying, Z.~Qian, H.~Zhou, H.~Xu, X.~Zhang, and S.~Li, ``From image to imuge:
  Immunized image generation,'' in \emph{Proceedings of the 29th ACM
  international conference on Multimedia}, 2021, pp. 1--9.

\bibitem{liu2021jpeg}
K.~Liu, D.~Chen, J.~Liao, W.~Zhang, H.~Zhou, J.~Zhang, W.~Zhou, and N.~Yu,
  ``Jpeg robust invertible grayscale,'' \emph{IEEE Transactions on
  Visualization and Computer Graphics}, 2021.

\bibitem{jia2021mbrs}
Z.~Jia, H.~Fang, and W.~Zhang, ``Mbrs: Enhancing robustness of dnn-based
  watermarking by mini-batch of real and simulated jpeg compression,'' in
  \emph{Proceedings of the 29th ACM International Conference on Multimedia},
  2021, pp. 41--49.

\bibitem{yu2015multi}
F.~Yu and V.~Koltun, ``Multi-scale context aggregation by dilated
  convolutions,'' \emph{arXiv preprint arXiv:1511.07122}, 2015.

\bibitem{kingma2014adam}
D.~P. Kingma and J.~Ba, ``Adam: A method for stochastic optimization,''
  \emph{arXiv preprint arXiv:1412.6980}, 2014.

\bibitem{cheng2004simple}
H.~Cheng and X.~Shi, ``A simple and effective histogram equalization approach
  to image enhancement,'' \emph{Digital signal processing}, vol.~14, no.~2, pp.
  158--170, 2004.

\bibitem{bruhn2005lucas}
A.~Bruhn, J.~Weickert, and C.~Schn{\"o}rr, ``Lucas/kanade meets horn/schunck:
  Combining local and global optic flow methods,'' \emph{International Journal
  of Computer Vision (IJCV)}, vol.~61, no.~3, pp. 211--231, 2005.

\bibitem{bay2006surf}
H.~Bay, T.~Tuytelaars, and L.~Van~Gool, ``Surf: Speeded up robust features,''
  in \emph{European Conference on Computer Vision (ECCV)}.\hskip 1em plus 0.5em
  minus 0.4em\relax Springer, 2006, pp. 404--417.

\bibitem{lowe1999object}
D.~G. Lowe, ``Object recognition from local scale-invariant features,'' in
  \emph{Proceedings of the IEEE International Conference on Computer Vision
  (ICCV)}, vol.~2.\hskip 1em plus 0.5em minus 0.4em\relax Ieee, 1999, pp.
  1150--1157.

\bibitem{wang2020cnn}
S.-Y. Wang, O.~Wang, R.~Zhang, A.~Owens, and A.~A. Efros, ``Cnn-generated
  images are surprisingly easy to spot... for now,'' in \emph{Proceedings of
  the IEEE/CVF Conference on Computer Vision and Pattern Recognition (CVPR)},
  2020, pp. 8695--8704.

\bibitem{lu2003fragile}
H.~Lu, R.~Shen, and F.-L. Chung, ``Fragile watermarking scheme for image
  authentication,'' \emph{Electronics Letters}, vol.~39, no.~12, pp. 898--900,
  2003.

\bibitem{he2006wavelet}
H.~He, J.~Zhang, and H.-M. Tai, ``A wavelet-based fragile watermarking scheme
  for secure image authentication,'' in \emph{International Workshop on Digital
  Watermarking}.\hskip 1em plus 0.5em minus 0.4em\relax Springer, 2006, pp.
  422--432.

\bibitem{zhang2011self}
X.~Zhang, S.~Wang, Z.~Qian, and G.~Feng, ``Self-embedding watermark with
  flexible restoration quality,'' \emph{Multimedia Tools and Applications},
  vol.~54, no.~2, pp. 385--395, 2011.

\bibitem{zhang2008fragile}
X.~Zhang and S.~Wang, ``Fragile watermarking with error-free restoration
  capability,'' \emph{IEEE Transactions on Multimedia}, vol.~10, no.~8, pp.
  1490--1499, 2008.

\bibitem{zhang2009fragile}
X.~Zhang, ``Fragile watermarking scheme using a hierarchical mechanism,''
  \emph{Signal processing}, vol.~89, no.~4, pp. 675--679, 2009.

\bibitem{korus2012efficient}
P.~Korus and A.~Dziech, ``Efficient method for content reconstruction with
  self-embedding,'' \emph{IEEE Transactions on Image Processing}, vol.~22,
  no.~3, pp. 1134--1147, 2012.

\bibitem{zhang2010reference}
X.~Zhang, Z.~Qian, and G.~Feng, ``Reference sharing mechanism for watermark
  self-embedding,'' \emph{IEEE Transactions on Image Processing (TIP)},
  vol.~20, no.~2, pp. 485--495, 2010.

\bibitem{zhang2011watermarking}
X.~Zhang, Z.~Qian, Y.~Ren, and G.~Feng, ``Watermarking with flexible
  self-recovery quality based on compressive sensing and compositive
  reconstruction,'' \emph{IEEE Transactions on Information Forensics and
  Security (TIFS)}, vol.~6, no.~4, pp. 1223--1232, 2011.

\bibitem{verdoliva2020media}
L.~Verdoliva, ``Media forensics and deepfakes: an overview,'' \emph{IEEE
  Journal of Selected Topics in Signal Processing}, vol.~14, no.~5, pp.
  910--932, 2020.

\bibitem{bi2019rru}
X.~Bi, Y.~Wei, B.~Xiao, and W.~Li, ``Rru-net: The ringed residual u-net for
  image splicing forgery detection,'' in \emph{Proceedings of the IEEE/CVF
  Conference on Computer Vision and Pattern Recognition Workshops (CVPRW)},
  2019, pp. 0--0.

\bibitem{wu2019mantra}
Y.~Wu, W.~AbdAlmageed, and P.~Natarajan, ``Mantra-net: Manipulation tracing
  network for detection and localization of image forgeries with anomalous
  features,'' in \emph{Proceedings of the IEEE/CVF Conference on Computer
  Vision and Pattern Recognition (CVPR)}, 2019, pp. 9543--9552.

\bibitem{chen2017self}
F.~Chen, H.~He, and Y.~Huo, ``Self-embedding watermarking scheme against jpeg
  compression with superior imperceptibility,'' \emph{Multimedia Tools and
  Applications}, vol.~76, no.~7, pp. 9681--9712, 2017.

\bibitem{preda2015watermarking}
R.~Preda and D.~Vizireanu, ``Watermarking-based image authentication robust to
  jpeg compression,'' \emph{Electronics Letters}, vol.~51, no.~23, pp.
  1873--1875, 2015.

\bibitem{tsai2008authentication}
M.-J. Tsai and C.-C. Chien, ``Authentication and recovery for wavelet-based
  semifragile watermarking,'' \emph{Optical Engineering}, vol.~47, no.~6, p.
  067005, 2008.

\bibitem{li2020face}
L.~Li, J.~Bao, T.~Zhang, H.~Yang, D.~Chen, F.~Wen, and B.~Guo, ``Face x-ray for
  more general face forgery detection,'' in \emph{Proceedings of the IEEE/CVF
  Conference on Computer Vision and Pattern Recognition (CVPR)}, 2020, pp.
  5001--5010.

\bibitem{bruna2011crop}
A.~R. Bruna, G.~Messina, and S.~Battiato, ``Crop detection through blocking
  artefacts analysis,'' in \emph{International Conference on Image Analysis and
  Processing}.\hskip 1em plus 0.5em minus 0.4em\relax Springer, 2011, pp.
  650--659.

\bibitem{fanfani2020vision}
M.~Fanfani, M.~Iuliani, F.~Bellavia, C.~Colombo, and A.~Piva, ``A vision-based
  fully automated approach to robust image cropping detection,'' \emph{Signal
  Processing: Image Communication}, vol.~80, p. 115629, 2020.

\bibitem{van2020dissecting}
B.~Van~Hoorick and C.~Vondrick, ``Dissecting image crops,'' \emph{arXiv
  preprint arXiv:2011.11831}, 2020.

\bibitem{goodfellow2014generative}
I.~J. Goodfellow, J.~Pouget-Abadie, M.~Mirza, B.~Xu, D.~Warde-Farley, S.~Ozair,
  A.~Courville, and Y.~Bengio, ``Generative adversarial networks,'' \emph{arXiv
  preprint arXiv:1406.2661}, 2014.

\bibitem{dinh2014nice}
L.~Dinh, D.~Krueger, and Y.~Bengio, ``Nice: Non-linear independent components
  estimation,'' \emph{arXiv preprint arXiv:1410.8516}, 2014.

\bibitem{dinh2016density}
L.~Dinh, J.~Sohl-Dickstein, and S.~Bengio, ``Density estimation using real
  nvp,'' \emph{arXiv preprint arXiv:1605.08803}, 2016.

\bibitem{kingma2018glow}
D.~P. Kingma and P.~Dhariwal, ``Glow: Generative flow with invertible 1x1
  convolutions,'' \emph{Advances in Neural Information Processing Systems
  (NIPS)}, vol.~31, 2018.

\bibitem{zhao2021invertible}
R.~Zhao, T.~Liu, J.~Xiao, D.~P. Lun, and K.-M. Lam, ``Invertible image
  decolorization,'' \emph{IEEE Transactions on Image Processing (TIP)},
  vol.~30, pp. 6081--6095, 2021.

\bibitem{xiao2020invertible}
M.~Xiao, S.~Zheng, C.~Liu, Y.~Wang, D.~He, G.~Ke, J.~Bian, Z.~Lin, and T.-Y.
  Liu, ``Invertible image rescaling,'' in \emph{European Conference on Computer
  Vision (ECCV)}.\hskip 1em plus 0.5em minus 0.4em\relax Springer, 2020, pp.
  126--144.

\bibitem{xing2021invertible}
Y.~Xing, Z.~Qian, and Q.~Chen, ``Invertible image signal processing,'' in
  \emph{Proceedings of the IEEE/CVF Conference on Computer Vision and Pattern
  Recognition (CVPR)}, 2021, pp. 6287--6296.

\bibitem{lu2021large}
S.-P. Lu, R.~Wang, T.~Zhong, and P.~L. Rosin, ``Large-capacity image
  steganography based on invertible neural networks,'' in \emph{Proceedings of
  the IEEE/CVF Conference on Computer Vision and Pattern Recognition (CVPR)},
  2021, pp. 10\,816--10\,825.

\bibitem{ILSVRC}
\BIBentryALTinterwordspacing
R.~O. Park~E, Liu~W, ``Ilsvrc-2017,'' 2017. [Online]. Available:
  \url{http://www.image-net.org/challenges/LSVRC/2017}
\BIBentrySTDinterwordspacing

\bibitem{karras2017progressive}
T.~Karras, T.~Aila, S.~Laine, and J.~Lehtinen, ``Progressive growing of gans
  for improved quality, stability, and variation,'' \emph{arXiv preprint
  arXiv:1710.10196}, 2017.

\bibitem{ying2021no}
Q.~Ying, X.~Hu, X.~Zhang, Z.~Qian, S.~Li, and X.~Zhang, ``Rwn: Robust
  watermarking network for image cropping localization,'' in \emph{2022 IEEE
  International Conference on Image Processing (ICIP)}.\hskip 1em plus 0.5em
  minus 0.4em\relax IEEE, 2022, pp. 301--305.

\bibitem{miyato2018spectral}
T.~Miyato, T.~Kataoka, M.~Koyama, and Y.~Yoshida, ``Spectral normalization for
  generative adversarial networks,'' \emph{arXiv preprint arXiv:1802.05957},
  2018.

\bibitem{iizuka2017globally}
S.~Iizuka, E.~Simo-Serra, and H.~Ishikawa, ``Globally and locally consistent
  image completion,'' \emph{ACM Transactions on Graphics (ToG)}, vol.~36,
  no.~4, pp. 1--14, 2017.

\bibitem{ulyanov2018deep}
D.~Ulyanov, A.~Vedaldi, and V.~Lempitsky, ``Deep image prior,'' in
  \emph{Proceedings of the IEEE/CVF Conference on Computer Vision and Pattern
  Recognition (CVPR)}, 2018, pp. 9446--9454.

\bibitem{otsu1979threshold}
N.~Otsu, ``A threshold selection method from gray-level histograms,''
  \emph{IEEE transactions on systems, man, and cybernetics}, vol.~9, no.~1, pp.
  62--66, 1979.

\bibitem{fridrich2007statistically}
J.~Fridrich, T.~Pevn{\`y}, and J.~Kodovsk{\`y}, ``Statistically undetectable
  jpeg steganography: dead ends challenges, and opportunities,'' in
  \emph{Proceedings of the 9th workshop on Multimedia \& security}, 2007, pp.
  3--14.

\bibitem{islam2020doa}
A.~Islam, C.~Long, A.~Basharat, and A.~Hoogs, ``Doa-gan: Dual-order attentive
  generative adversarial network for image copy-move forgery detection and
  localization,'' in \emph{Proceedings of the IEEE/CVF Conference on Computer
  Vision and Pattern Recognition (CVPR)}, 2020, pp. 4676--4685.

\bibitem{schaefer2003ucid}
G.~Schaefer and M.~Stich, ``Ucid: An uncompressed color image database,'' in
  \emph{Storage and Retrieval Methods and Applications for Multimedia 2004},
  vol. 5307.\hskip 1em plus 0.5em minus 0.4em\relax International Society for
  Optics and Photonics, 2003, pp. 472--480.

\bibitem{agustsson2017ntire}
E.~Agustsson and R.~Timofte, ``Ntire 2017 challenge on single image
  super-resolution: Dataset and study,'' in \emph{Proceedings of the IEEE/CVF
  Conference on Computer Vision and Pattern Recognition Workshops (CVPRW)},
  2017, pp. 126--135.

\bibitem{CASIA}
J.~Dong, W.~Wang, and T.~Tan, ``Casia image tampering detection evaluation
  database,'' in \emph{2013 IEEE China Summit and International Conference on
  Signal and Information Processing}.\hskip 1em plus 0.5em minus 0.4em\relax
  IEEE, 2013, pp. 422--426.

\bibitem{DEFACTO}
G.~Mahfoudi, B.~Tajini, F.~Retraint, F.~Morain-Nicolier, J.~L. Dugelay, and
  P.~Marc, ``Defacto: Image and face manipulation dataset,'' in \emph{2019 27th
  European Signal Processing Conference (EUSIPCO)}.\hskip 1em plus 0.5em minus
  0.4em\relax IEEE, 2019, pp. 1--5.

\bibitem{he2017mask}
K.~He, G.~Gkioxari, P.~Doll{\'a}r, and R.~Girshick, ``Mask r-cnn,'' in
  \emph{Proceedings of the IEEE International Conference on Computer Vision
  (ICCV)}, 2017, pp. 2961--2969.

\bibitem{ren2015faster}
S.~Ren, K.~He, R.~Girshick, and J.~Sun, ``Faster r-cnn: Towards real-time
  object detection with region proposal networks,'' \emph{Advances in Neural
  Information Processing Systems (NIPS)}, vol.~28, pp. 91--99, 2015.

\bibitem{liu2018large}
Z.~Liu, P.~Luo, X.~Wang, and X.~Tang, ``Large-scale celebfaces attributes
  (celeba) dataset,'' \emph{Retrieved August}, vol.~15, no. 2018, p.~11, 2018.

\bibitem{zhou2017places}
B.~Zhou, A.~Lapedriza, A.~Khosla, A.~Oliva, and A.~Torralba, ``Places: A 10
  million image database for scene recognition,'' \emph{IEEE Transactions on
  Pattern Analysis and Machine Intelligence (TPAMI)}, vol.~40, no.~6, pp.
  1452--1464, 2017.

\bibitem{grathwohl2018ffjord}
W.~Grathwohl, R.~T. Chen, J.~Bettencourt, I.~Sutskever, and D.~Duvenaud,
  ``Ffjord: Free-form continuous dynamics for scalable reversible generative
  models,'' \emph{arXiv preprint arXiv:1810.01367}, 2018.

\bibitem{ardizzone2018analyzing}
L.~Ardizzone, J.~Kruse, S.~Wirkert, D.~Rahner, E.~W. Pellegrini, R.~S. Klessen,
  L.~Maier-Hein, C.~Rother, and U.~K{\"o}the, ``Analyzing inverse problems with
  invertible neural networks,'' \emph{arXiv preprint arXiv:1808.04730}, 2018.

\bibitem{zhu2017unpaired}
J.-Y. Zhu, T.~Park, P.~Isola, and A.~A. Efros, ``Unpaired image-to-image
  translation using cycle-consistent adversarial networks,'' in
  \emph{Proceedings of the IEEE International Conference on Computer Vision
  (ICCV)}, 2017, pp. 2223--2232.

\bibitem{pevny2010using}
T.~Pevn{\`y}, T.~Filler, and P.~Bas, ``Using high-dimensional image models to
  perform highly undetectable steganography,'' in \emph{International Workshop
  on Information Hiding}.\hskip 1em plus 0.5em minus 0.4em\relax Springer,
  2010, pp. 161--177.

\bibitem{cheddad2010digital}
A.~Cheddad, J.~Condell, K.~Curran, and P.~Mc~Kevitt, ``Digital image
  steganography: Survey and analysis of current methods,'' \emph{Signal
  processing}, vol.~90, no.~3, pp. 727--752, 2010.

\bibitem{cox2007digital}
I.~Cox, M.~Miller, J.~Bloom, J.~Fridrich, and T.~Kalker, \emph{Digital
  watermarking and steganography}.\hskip 1em plus 0.5em minus 0.4em\relax
  Morgan kaufmann, 2007.

\bibitem{hayes2017generating}
J.~Hayes and G.~Danezis, ``Generating steganographic images via adversarial
  training,'' \emph{arXiv preprint arXiv:1703.00371}, 2017.

\bibitem{tang2017automatic}
W.~Tang, S.~Tan, B.~Li, and J.~Huang, ``Automatic steganographic distortion
  learning using a generative adversarial network,'' \emph{IEEE Signal
  Processing Letters}, vol.~24, no.~10, pp. 1547--1551, 2017.

\bibitem{Wu_2018}
\BIBentryALTinterwordspacing
P.~Wu, Y.~Yang, and X.~Li, ``Stegnet: Mega image steganography capacity with
  deep convolutional network,'' \emph{Future Internet}, vol.~10, no.~6, p.~54,
  Jun 2018. [Online]. Available: \url{http://dx.doi.org/10.3390/fi10060054}
\BIBentrySTDinterwordspacing

\bibitem{2019SteganoGAN}
K.~A. Zhang, A.~Cuesta-Infante, and K.~Veeramachaneni, ``Steganogan: Pushing
  the limits of image steganography,'' 2019.

\bibitem{wengrowski2019light}
E.~Wengrowski and K.~Dana, ``Light field messaging with deep photographic
  steganography,'' in \emph{Proceedings of the IEEE/CVF Conference on Computer
  Vision and Pattern Recognition (CVPR)}, 2019, pp. 1515--1524.

\bibitem{park2021image}
C.~W. Park, Y.~H. Moon, and I.~K. Eom, ``Image tampering localization using
  demosaicing patterns and singular value based prediction residue,''
  \emph{IEEE Access}, vol.~9, pp. 91\,921--91\,933, 2021.

\bibitem{le2019improved}
N.~Le and F.~Retraint, ``An improved algorithm for digital image authentication
  and forgery localization using demosaicing artifacts,'' \emph{IEEE Access},
  vol.~7, pp. 125\,038--125\,053, 2019.

\bibitem{choi2019evaluating}
J.-H. Choi, H.~Zhang, J.-H. Kim, C.-J. Hsieh, and J.-S. Lee, ``Evaluating
  robustness of deep image super-resolution against adversarial attacks,'' in
  \emph{Proceedings of the IEEE/CVF International Conference on Computer Vision
  (ICCV)}, 2019, pp. 303--311.

\bibitem{neekhara2021adversarial}
P.~Neekhara, B.~Dolhansky, J.~Bitton, and C.~C. Ferrer, ``Adversarial threats
  to deepfake detection: A practical perspective,'' in \emph{Proceedings of the
  IEEE/CVF Conference on Computer Vision and Pattern Recognition (CVPR)}, 2021,
  pp. 923--932.

\bibitem{karras2019style}
T.~Karras, S.~Laine, and T.~Aila, ``A style-based generator architecture for
  generative adversarial networks,'' in \emph{Proceedings of the IEEE/CVF
  Conference on Computer Vision and Pattern Recognition (CVPR)}, 2019, pp.
  4401--4410.

\bibitem{flickr1024}
Y.~Wang, L.~Wang, J.~Yang, W.~An, and Y.~Guo, ``Flickr1024: A large-scale
  dataset for stereo image super-resolution,'' in \emph{The IEEE International
  Conference on Computer Vision Workshops (ICCVW)}, Oct 2019, pp. 3852--3857.

\bibitem{schonfeld2020u}
E.~Schonfeld, B.~Schiele, and A.~Khoreva, ``A u-net based discriminator for
  generative adversarial networks,'' in \emph{Proceedings of the IEEE/CVF
  Conference on Computer Vision and Pattern Recognition (CVPR)}, 2020, pp.
  8207--8216.

\bibitem{PASSRnet}
L.~Wang, Y.~Wang, Z.~Liang, Z.~Lin, J.~Yang, W.~An, and Y.~Guo, ``Learning
  parallax attention for stereo image super-resolution,'' in \emph{Proceedings
  of the IEEE/CVF Conference on Computer Vision and Pattern Recognition
  (CVPR)}, 2019, p. 12250–12259.

\end{thebibliography}

\begin{IEEEbiography}[{\includegraphics[width=1in,height=1.25in,clip,keepaspectratio]{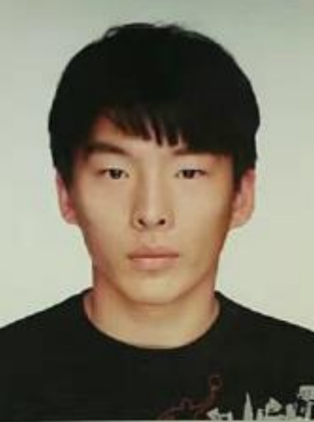}}]{Qichao Ying}
received the B.S. and M.S. degree from the School of Communication and Information Engineering, Shanghai University, China, respectively in 2017 and 2020. He is currently a doctoral candidate in the School of Computer Science, Fudan University, China. His research interests include multimedia forensics, data hiding and fake news detection.
\end{IEEEbiography}

\begin{IEEEbiography}[{\includegraphics[width=1in,height=1.25in,clip,keepaspectratio]{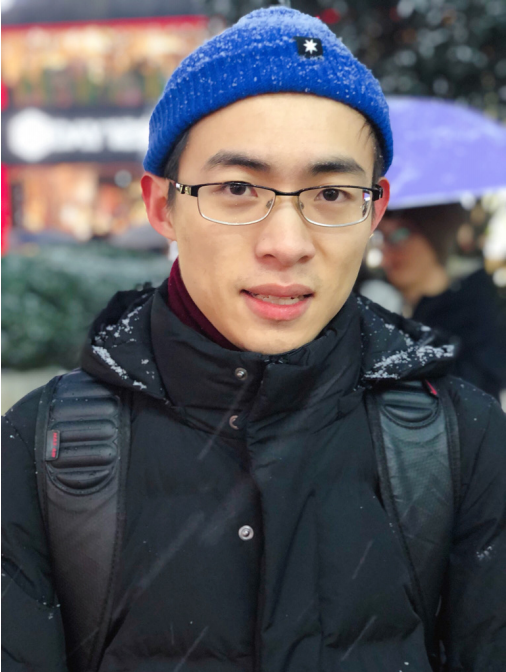}}]{Hang Zhou}
received his B.S. degree in 2015 from Shanghai University and a Ph.D. degree in 2020 from the University of Science and Technology of China. Currently, he is a postdoctoral researcher at Simon Fraser University. His research interests include computer graphics, multimedia security and deep learning.
\end{IEEEbiography}

\begin{IEEEbiography}[{\includegraphics[width=1in,height=1.25in,clip,keepaspectratio]{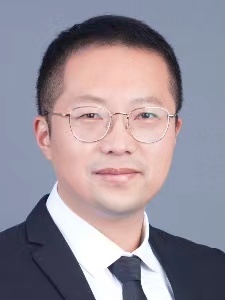}}]{Zhenxing Qian}
received the B.S. and Ph.D. degrees from the University of Science and Technology of China (USTC), in 2003 and 2007, respectively. He is currently a Professor with the School of Computer Science, Fudan University. He has published over 100 peer-reviewed papers on international journals and conferences. His research interests include information hiding, image processing, and multimedia security.

\end{IEEEbiography}

\begin{IEEEbiography}[{\includegraphics[width=1in,height=1.25in,clip,keepaspectratio]{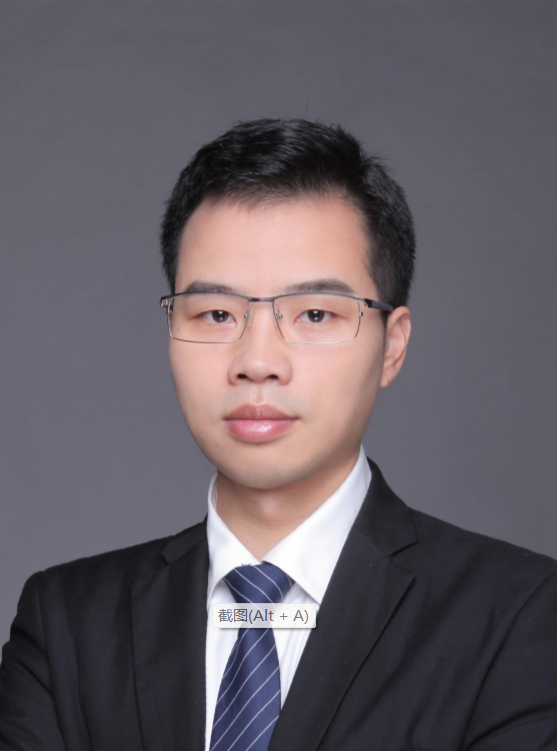}}]{Sheng Li}
received the Ph.D. degree at the School of Electrical and Electronic Engineering, Nanyang Technological University, Singapore, in 2013. From 2013 to 2016, he was a research fellow in Rapid Rich Object Search (ROSE) Lab, Nanyang Technological University. He is currently an Associate Professor with the School of Computer Science, Fudan University, China. His research interests include biometric template protection, pattern recognition, multimedia forensics and security. He is the recipient of the IEEE WIFS Best Student Paper Silver Award.

\end{IEEEbiography}

\begin{IEEEbiography}[{\includegraphics[width=1in,height=1.25in,clip,keepaspectratio]{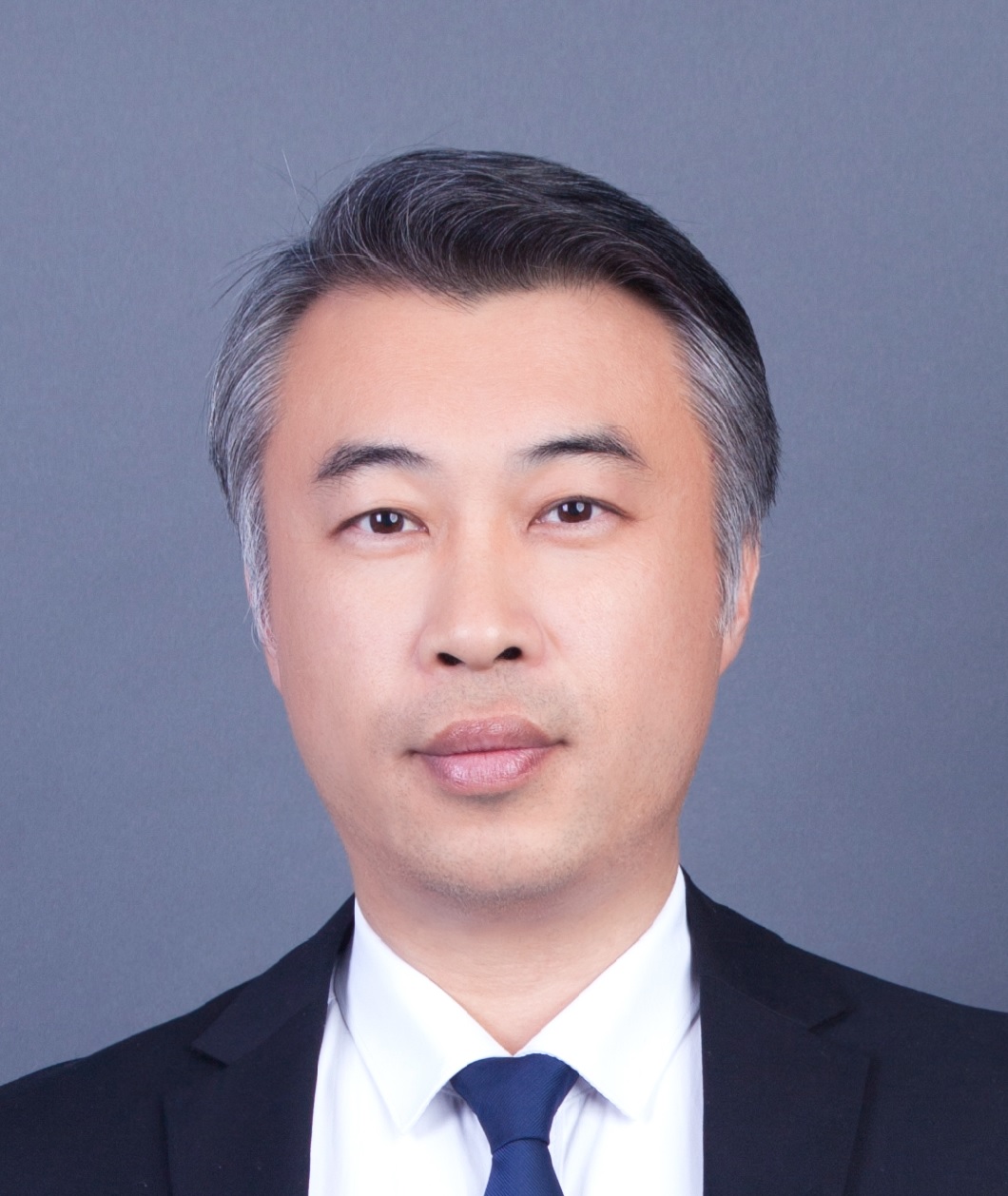}}]{Xinpeng Zhang}
received the B.S. degree in computational mathematics from Jilin University, China, in 1995, and the M.E. and Ph.D. degrees in communication and information system from Shanghai University, China, in 2001 and 2004, respectively, where he has been with the faculty of the School of Communication and Information Engineering, since 2004, and is currently a Professor. His research interests include information hiding, image processing, and digital forensics. He has published over 200 papers in these areas.
\end{IEEEbiography}

\end{document}